\documentclass[12pt,semidraft,epsfig]{article}
\usepackage{amsmath}
\usepackage{graphicx}

\usepackage{slashed}

\def\s#1{\setbox0=\hbox{$#1$}%
\rlap{\ifdim\wd0>.7em\kern.22\wd0\else\kern.1\wd0\fi /}#1}

\newcommand{\beq}{\begin{equation}}
\newcommand{\eeq}{\end{equation}}
\newcommand{\bea}{\begin{eqnarray}}
\newcommand{\eea}{\end{eqnarray}}

\newcommand{\tto}{\!\to\!}

\newcommand{\gsim}{\lower.7ex\hbox{$
\;\stackrel{\textstyle>}{\sim}\;$}}
\newcommand{\lsim}{\lower.7ex\hbox{$
\;\stackrel{\textstyle<}{\sim}\;$}}

\renewcommand{\Im}{{\rm Im}\,}

\newcommand{\bibit}[1]{\bibitem{#1}}

\newcommand{\aver}[1]{\langle #1\rangle}

\newcommand{\mhad}{\mu_{\rm hadr}}

\newcommand{\GeV}{\,\mbox{GeV}}
\newcommand{\MeV}{\,\mbox{MeV}}
\newcommand{\matel}[3]{\langle #1|#2|#3\rangle}

\newcommand{\msp}[1]{\mbox{\hspace*{#1mm}~}}

\begin{document}
\begin{titlepage}
\begin{flushright}
SI-HEP-2012-06 \\[0.2cm]

\end{flushright}

\vspace{1.2cm}
\begin{center}
{\Large\bf 
Charm CP Violation and the 
Electric Dipole \\[10pt] Moments from the Charm Scale}
\end{center}

\vspace{0.5cm}
\begin{center}
{\large \sc Thomas Mannel\,  and\, Nikolai Uraltsev} \\[0.4cm]
{\sf Theoretische Elementarteilchenphysik, Naturwiss.-Techn.\ Fakult\"at, \\[0.1cm]
Universit\"at Siegen, 57068 Siegen, Germany}
\end{center}

\vspace{0.8cm}
\begin{abstract}

\noindent
The reported CP asymmetry in $D\tto K^+K^-/\pi^+\pi^-$ is argued
to be too large to naturally fit the SM. If so, a new source of CP violation is
implied in the $\Delta C\!=\!1$ sector with a milliweak strength. CP-odd
interactions in the flavor-diagonal sector are strongly constrained by the
EDMs placing severe limitations on the underlying theory. While the largest effects
usually come from the New Physics energy scale, they are strongly 
model-dependent. Yet the interference
of the CP-odd forces manifested in $D$ decays with the conventional CP-even
$\Delta C\!=\!1$ weak interaction generates at the charm scale 
a background level.  It has 
been argued that the $d_n$ in the SM is largely generated via such an 
interference, with mild KM-specific additional suppression.
The reported CP asymmetry is expected to generate $d_n$ of
$30$ to $100$ times larger than in the SM, or even higher in certain model yet
not quite natural examples. In the SM the charm-induced 
loop-less $|d_n|$ is expected around $10^{-31}\,\mbox{e$\cdot$cm}$.
On the technical side, we present a compact Ward-identity--based derivation of the
induced scalar pion-nucleon coupling in the presence of the CP-odd
interactions, which appears once the latter include the right-handed light
quarks.

\end{abstract}

\thispagestyle{empty}

\addtocounter{page}{-1}

\end{titlepage}


\pagenumbering{arabic}
\section{Introduction}
Recent hints at possible direct CP violation in singly Cabibbo-suppressed $D$
meson decays have caused some excitement, since they may be the first direct
indication for physics beyond the Standard Model (SM). 
The LHCb
collaboration observes a difference of time-integrated CP asymmetries \cite{lhcb}
\begin{equation}
\Delta a_{\rm CP} = a_{\rm CP} (D^0 \to K^+ K^-) - a_{\rm CP} (D^0 \to \pi^+ \pi^-) 
= - (0.82 \pm 0.21 \pm 0.11)\% ,
\label{t1}
\end{equation}  
a result preliminarily confirmed by CDF \cite{cdf}:
\begin{equation}
\Delta a_{\rm CP} = - (0.62 \pm 0.21 \pm 0.10)\% .
\label{t2}
\end{equation}  
The CP asymmetry is defined according to 
\begin{equation} 
a_{\rm CP} (D^0 \to f) = \frac{\Gamma (D^0 \to f) - 
\Gamma (\overline{D}^0 \to \overline{f})}
  {\Gamma (D^0 \to f) + \Gamma (\overline{D}^0 \to \overline{f})}
\end{equation} 
 Since the both final states are positive CP eigenstates in strong 
interactions, $\Delta a_{\rm CP}$ evidently roots in a direct CP asymmetry.  

As argued below, an effect of this magnitude solely in the framework of 
the SM may not be
rigorously excluded, yet it would require a strong enhancement of certain 
decay matrix elements.
Such a loophole definitely deserves further scrutiny.
In this paper we analyze immediate consequences of the assumption 
that the reported effect is due to a
new source of CP violation, beyond the CKM mechanism.  

In the present paper we focus on the impact of new CP-odd forces on the
electric dipole moments, in particular on one of the neutron, $d_n$.  The observation
at LHCb and CDF assumes CP violation in $|\Delta C| \!=\! 1$ amplitudes. A more
general model should embed this into a full flavor framework, hopefully
highlighting a certain underlying 
symmetry. It would allow one to obtain predictions for
other sensitive processes as well, in particular for $B$ and $K$ decays, where
some tensions have also been noted in certain cases.

We do not attempt to elaborate such a framework here and rather moderate the
ambitions to noting that a flavor-diagonal CP violation of the size reported
in the decays $D^0 \to K^+K^-$ and $D^0 \to \pi^+\pi^-$ is incompatible with
the current limits on the electric dipole moment of the neutron. At the same
time, we find that if the new CP-odd forces show up at low energies only in
$|\Delta C|\!=\!1$ interactions, the neutron EDM is still well below the
current limit, although it should be significantly enhanced, by more than 
an order of magnitude, compared to the SM.

In the following section we briefly review the CP violation in $D^0 \tto K^+K^-$
and $D^0 \tto \pi^+\pi^-$ within the SM and introduce new $|\Delta C|\!=\! 1$
interactions as a source of the enhanced CP violation. In Sect.~3 we examine 
$d_n$ in the SM and describe the loop-free mechanism to generate it at the
charm scale.  The elaborated estimates of the associated nucleon matrix
elements indicate that it yields $d_n$ around 
$10^{-31\,}\textrm{e}\!\cdot\!\textrm{cm}$ and  may well constitute the
principal contribution in the SM. 
The same analysis is then adapted to new BSM-mediated CP-odd amplitudes 
to estimate the corresponding effect on the neutron EDM. 
Section~4 summarizes the study. Appendix derives the induced CP-odd scalar
pion-nucleon coupling by generalizing the Goldberger-Treiman
relation and applying in QCD the current algebra technique.

\section{\boldmath CP Violation in $D^0 \to K^+K^-$ and $D^0 \to \pi^+\pi^-$ }
\label{CPD}
\subsection{Charm CP Violation in the Standard Model}
\label{cpvsm}

For the charm decays considered hereafter we have the effective weak interaction  
\bea
\nonumber 
{\cal L}\msp{-4} &=&\msp{-4}
-\frac{G_F}{\sqrt{2}}\left[\frac{V_{cs}V^*_{us}\!-\!V_{cd}V^*_{ud}}{2} 
\left([\bar{c}\Gamma^\mu u][\bar{s}\Gamma_\mu s] -
[\bar{c}\Gamma^\mu u][\bar{d}\Gamma_\mu d]
\right) - \right. \\
\msp{-4} & &\msp{-4} \left. \msp{12}
\mbox{$\frac{1}{2}$}\, V_{cb}V^*_{ub}
\left([\bar{c}\Gamma^\mu u][\bar{s}\Gamma_\mu s] +
[\bar{c}\Gamma^\mu u][\bar{d}\Gamma_\mu d]
- 2[\bar{c}\Gamma^\mu u][\bar{b}\Gamma_\mu b ] \right), 
\rule[15pt]{1pt}{0pt}
\right] + \mbox{H.c.}
\label{404} \\
\nonumber
\msp{-4} &= &\msp{-4} 
-\frac{G_F}{\sqrt{2}} \sin{\theta_C} \cos{\theta_C}
\left[ o_1 - \mbox{$\frac{1}{2}$}r_{\rm SM} e^{-i \gamma} o_2 \right] + 
\mbox{H.c.},\rule[-15pt]{10pt}{0pt}\\
\nonumber
&&\msp{-14} [\bar{c}\Gamma^\mu u][\bar{q}\Gamma_\mu q] \equiv 
(\bar{c}\gamma^\mu(1\!-\!\gamma_5)q) \,(\bar{q}\gamma_\mu(1\!-\!\gamma_5)u),
\qquad \Gamma_\alpha = \gamma_\alpha(1\!-\!\gamma_5), 
\eea
where we have used the CKM unitarity, and color indices are assumed to be
contracted within parentheses. The phase $\gamma$ is practically equal to the
corresponding angle of the Unitarity Triangle, while  
\begin{equation}
r_{\rm SM} = \left| \frac{V_{cb}^* V_{ub}}{V_{cs}^* V_{us}} \right| \simeq 7.5
\times 10^{-4} .
\end{equation} 
CP violation in the SM is quantified by 
the imaginary part of the invariant product of four CKM mixing elements
describing the relative phase between the coefficients for $o_1$ and $o_2$,
\begin{equation}
\Delta  =   {\rm Im} \, V_{cs}^* V_{us} V_{cd} V_{ud}^* = 
{\rm Im} \, (V_{cb}^* V_{cd})^* (V_{ub}^*V_{ud}) ,  
\label{402}
\end{equation} 
numerically $\Delta \!\simeq\! 3.3 \cdot 10^{-5}$.

The operators $o_1$ and $o_2$
are a U-spin triplet and a U-spin singlet, respectively.
Their interference induces the CP violation in the SM in the $\Delta C\!=\!1$
sector. In what follows we discard possible CP violation in $\bar D\!-\!D$
mixing since it drops out from the asymmetry difference $\Delta a_{\rm CP}$.

The decay amplitudes in $D^0 \to f$ with  $f\!=\!\bar f$
are given by ($f \!=\! K^+ K^-$ or $f \!=\! \pi^+ \pi^-$)
\begin{equation} 
A (D^0 \to f) =  -i \frac{G_F}{\sqrt{2}} \sin \theta_C \cos \theta_C 
\left[ m_1^{(f)} - \mbox{$\frac{1}{2}$}r_{\rm SM} e^{-i \gamma} m_2^{(f)} \right] 
\label{amp}
\end{equation} 
where $m_i^{(f)} \!=\! \matel{f}{o_i}{D^0}$ are the reduced amplitudes, in
general complex due to the strong interaction in the final state. The
corresponding phases are generically referred to as $\delta_i^{(f)}$; they are
equal for the decays of $D$ and $\bar D$.
Since the $o_1$ amplitudes strongly dominate, $r_{\rm SM}\!\ll \! 1$, 
the CP asymmetry takes a simple form
\begin{equation}
a_{\rm CP} (D^0 \tto f) =  -r_{\rm SM}\sin{\gamma}   \left| \frac{m_2^{(f)}}{m_1^{(f)}}
\right| \, 
\sin{\delta_{21}^{(f)}}, \qquad r_{\rm SM}\sin{\gamma}\simeq 
\frac{\Delta}{\sin^2{\theta_C}\cos^2{\theta_C}} \!\simeq\! 0.70\cdot 10^{-3},
\label{acp}
\end{equation}
where $\delta_{21}^{(f)}\!=\!\delta_2^{(f)}\!-\!\delta_1^{(f)}$ is the phase 
difference between the two hadronic matrix elements. 

The two final states  $K^+ K^-$ and $\pi^+ \pi^-$ are components of the same
U-spin triplet. Therefore in the $SU(3)$ limit $\delta_{21}^{\pi^+ \pi^-} \!=\! 
 \delta_{21}^{K^+ K^-} \!+\! \pi $ would hold and 
\begin{equation}
| \Delta a_{\rm CP} |  \simeq 2 |a_{CP} (D^0 \tto K^+ K^- ) |\simeq 2 
|a_{CP} (D^0 \tto \pi^+ \pi^- ) |.
\end{equation} 
However, U-spin symmetry is significantly violated;  one concludes from the 
decay rates 
\begin{equation}
\Gamma (D^0 \tto P\bar{P}) = \frac{G_F^2}{32 \pi M_D}   \sin^2{\theta_C} \cos^2{\theta_C}
\sqrt{1\!-\!\frac{4 M_P^2}{M_D^2}}\: |m_1^{(P\bar{P})} |^2
\end{equation} 
that  
\begin{equation}
|m_1^{K^+K^-} |\simeq 0.456 \, {\rm GeV}^3, \qquad 
|m_1^{\pi^+\pi^-} |\simeq 0.252 \, {\rm GeV}^3.
\label{432}
\end{equation} 
We expect even larger potential $SU(3)$ breaking in the phases of
the amplitudes. 

The values in Eq.~(\ref{432}) reasonably agree with the simplest factorization
estimate
\beq
m_1^{K^+K^-}\!\simeq\! i f_+^{D\tto K}(M_K^2)f_K (M_D^2\!-\!M_K^2), \quad 
m_1^{\pi^+\pi^-}\!\simeq\! - i f_+^{D\tto \pi}(M_\pi^2)f_\pi(M_D^2\!-\!M_\pi^2)
\label{434}
\eeq
(the straightforward color renormalization factors have been omitted from the 
full expression). It even yields the right scale for the $SU(3)$-breaking \cite{ru},
although the literal ratio of the amplitudes tends to fall short of $1.81$ in 
Eq.~(\ref{432}).

The matrix elements of $o_2$ determining the amplitudes $m_2^{K^+K^-}$ and
$m_2^{\pi^+\pi^-}$ are not directly known. If estimated using factorization,
one evidently obtains the values close to $m_1$ in Eqs.~(\ref{434}), (\ref{432}),
with the additional minus sign for $m_2^{\pi^+\pi^-}$. However, the
conventional factorization accounts only for the valence contributions. In a
valence approximation, on the other hand, the same term in both $o_1$ and
$o_2$ -- with $s$-quarks for $D^0\tto K^+K^-$ and with $d$-quarks for 
$D^0\tto \pi^+\pi^-$, respectively -- contribute. Consequently, no CP-asymmetry
is generated in a valence approximation: the two strong amplitudes come from  
the same underlying operators and their strong phases coincide. 

Strictly speaking, any valence approximation should only be applied to the
operators normalized at a low scale. The evolution of the operators $o_1$ and
$o_2$  above $m_b$ is identical, yet generates additional
terms for $o_2$ below it due to Penguin diagrams \cite{penguins}. These in general have
different strong phases. However, the new operators come with small
loop-induced coefficients, while we do not expect their matrix elements to be
enhanced. Therefore, we neglect these effects.

As the starting point we assume that the magnitudes of $m_2^{K^+K^-}$ and
$m_2^{\pi^+\pi^-}$ may be approximated using factorization, yet allow for 
arbitrary FSI phases relative to $m_1$. This amounts to having the ratio of
the amplitudes in Eq.~(\ref{acp}) about unity. We then end up with
\beq
|a_{\rm CP}(D^0\tto K^+K^-)| \!\approx\! 0.7\cdot 10^{-3}
|\sin{\delta_{21}^{K^+K^-}}|, \quad 
|a_{\rm CP}(D^0\tto \pi^+\pi^-)| \!\approx\! 0.7\cdot 10^{-3}
|\sin{\delta_{21}^{\pi^+ \pi^-}}|,
\label{natur}
\eeq
and the sign of the two asymmetries may naturally be opposite. Therefore, the
expected scale for $|\Delta a_{\rm CP}|$ in the SM is a few times $10^{-4}$ up
to $1.5\cdot 10^{-3}$ -- provided the both FSI phase shifts 
are optimal. This is still about
five times smaller than what is reported by LHCb.

Accommodating the central value in Eq.~(\ref{t1}) within the SM thus implies at least
a five-fold enhancement of the U-spin singlet amplitude mediated by $o_2$, or even a
ten-fold if the asymmetry is dominated by one of the two modes
and/or the strong phase shifts are not optimal. Moreover, this must happen
for a non-valence part of the amplitude.

Although at the moment the possibility of a
sufficiently strong enhancement of the $U$-scalar amplitude in the $D$
decay within conventional QCD dynamics cannot be rigorously ruled out, 
we view this possibility as
contrived. The confirmation of the asymmetry $\Delta a_{\rm CP}$ 
at the currently observed level, 
in particular studying its share among the two
channels, would be a strong evidence for the new CP-violating dynamics in the
charm sector. At the same time, the strength of this conclusion crucially 
depends on the actual amount of the excess over our expectations. An eventual
value around or somewhat below $-0.3\%$, while still not smoothly 
accommodated in the SM, {\it per se} would make the case for new sources of
CP-violation in $D$ decays significantly weaker. 

\subsection{Charm CP Violation through New Physics}
\label{cpdbsm}

In what follows we adopt the assumption that the reported CP asymmetry roots
in new CP-odd interactions. Within this hypothesis we will not attempt to
stretch the uncertainties due to the QCD interaction to as strong extent and
rather apply an educated judgment elaborated in weak decays so far; we then
examine the consequences for the electric dipole moments. Neither we focus 
on the extreme values of parameters maximizing the CP
asymmetry. Consequently, we will gauge our expectations on an assumption that,
speaking generally, the new source of CP violation produces an {\sl order of
  magnitude} stronger CP-odd amplitude in $D \to K^+K^-$ or  in $D \to
\pi^+\pi^-$ decays than in the SM. 

Turning to NP, we make a relatively safe assumption that the New
Physics-induced amplitude is small compared to the SM one $m_1$; 
this is obvious for the CP-odd NP part, and is applied also to its CP-even
component. Then the asymmetry is given by the sum of the
pure NP-induced asymmetry and the SM one, for either channels.  Keeping in
mind the conclusions of Sect.~\ref{cpvsm} we neglect the SM 
contribution altogether, and have
\beq
a_{\rm CP} (D^0 \tto f) =  - 2\left| \frac{\Im g_{\rm NP}\: 
m_{\rm NP}^{(f)}}{m_1^{(f)}} \right| \, 
\sin{\delta_{\rm NP}^{(f)}} \, ,
\label{220}
\eeq
where  $g_{\rm NP}\, m_{\rm NP}^{(f)}$ denotes the New Physics amplitude (in units of 
$G_F \sin{\theta_C} \cos{\theta_C} / \sqrt{2}$) and $\delta_{NP}^{(f)}$ is its
strong phase relative to $m_1^{(f)}$ of the SM.\footnote{We assume $\Im
g_{\rm NP}\!>\!0$ and include the possible sign into  $\delta_{NP}^{(f)}$. Let
us also clarify that the phase convention required for CP conjugation is defined 
in such a way that the $a_1$ amplitude in the SM is CP-even.}

The value of the new couplings  $g_{\rm NP}$ depends on the convention chosen
to parameterize the BSM amplitudes (we tacitly anticipate using effective
local operators to describe them). If we assume a `natural' normalization
of the operators 
where $|m_{NP}^{(f)}|\!\approx \! |m_1^{(f)}|$ holds for the reduced
amplitudes, we arrive at a ballpark estimate for the CP-odd coupling:
\beq
|\Im g_{\rm NP}| \sim  (2\,\div \,5)\cdot 10^{-3},
\label{222}
\eeq
allowing for the generic unsuppressed strong phase differences as the
educated guess about QCD dynamics in charm. The new CP-odd 
forces must therefore be of a `milliweak'
strength, according to the venerable terminology in CP violation. 
Their strength is in general about $10$ times larger than what one 
estimates in the SM, cf. $r_{\rm SM} \sin{\gamma}$ in Eq.~(\ref{acp}).
Specifically, to accommodate the `direct' CP-asymmetry Eq.~(\ref{t1}) one needs 
\beq
0.55\frac{\Im g_{\rm NP}\: m_{\rm NP}^{K^+K^-}}{10^{-3}\GeV^3} 
\sin{\delta_{\rm NP}^{K^+K^-}} 
+\frac{\Im g_{\rm NP}\: m_{\rm NP}^{\pi^+\pi^-}}{10^{-3}\GeV^3}
\sin{\delta_{\rm NP}^{\pi^+\pi^-}} = \frac{-\Delta a_{\rm CP}}{8.2\cdot 10^{-3}}.
\label{514}
\eeq

Proceeding to the induced CP-odd effects in other observables requires specifying
the nature of new interactions. Below the charm scale we have the realm of
light hadrons including flavor-diagonal processes with stable hadrons and the
decays of strange particles. CP-odd effects there are highly constrained;
therefore we discard these, and relegate new sources to heavy particles. Their
effect at low energies is described by local operators classified over the
canonic dimension, with the lowest-dimension potentially dominating. 

A flavor-diagonal CP violation (for instance, the induced QCD $\theta$-term
unless it is offset by a Peccei-Quinn--type mechanism)
with a coupling of the size commensurate with Eq.~(\ref{222}) is by far excluded by
electric dipole moments, in particular of the neutron. This likewise applies to
the four-quark operators -- they would generate $d_n$ in the ball park
of $10^{-22}\,\mbox{e$\cdot$cm}$. Consequently, the 
flavor structure of the new CP-odd interaction must have vanishing 
flavor-diagonal components in the light sector. This property must replicate
itself at the loop level, which strongly suggests it to apply to the heavy
flavors alike. We do not analyze here the consequences of this requirement for
various classes of the BSM models, but rather note that this may be a hint at an
antisymmetric in generations structure of the underlying flavor dynamics. Yet
even when postulating such a property, nontrivial constraints may 
follow at the loop
level in view of the large gap between the scale of the potential effect 
on the EDMs and of 
their experimental limits. The loop-induced effects and the related
renormalization of the effective low-energy operators may strongly depend on a
particular class of models, 
see, e.g.\ Refs.~\cite{gino,gian}; this lies outside the scope of our analysis.

We therefore concentrate on the most direct consequences of the presence of
new $\Delta C\!=\!1$ CP-odd amplitudes and describe them by the effective
operators of dimension $5$ and $6$.
Most of them are four-quark operators. This appears representative enough.  
The reason, as argued below, is that
a significant -- and probably the dominant -- piece of $d_n$ in the Standard Model
likewise originates from the same underlying effect: the interference of the
$CP$-odd and $CP$-even weak $|\Delta C|\!=\!1$ amplitudes in the nucleon. The SM
bears only mild additional model-specific suppressions; these may, or may
not be vitiated by the new CP-odd $|\Delta C|\!=\!1$ interaction, depending on
particular details. Consequently,
we typically obtain a $30$- to $100$-fold enhancement of $d_n$ compared to the
SM.

The number of appropriate $D\!=\!6$ operators (they can be both scalar or
pseudoscalar) is quite large since they may differ in the chiral, color and
light-flavor content.  We first note that the scalar operators do not
affect either of the $D$ decays in question, yet they do
generate $d_n$. Therefore, $d_n$ could have been further
enhanced if the scalar NP operators dominate. This
possibility can be effectively eliminated experimentally by studying the
similar CP-asymmetries including the parity-even final states in decays of $D$
mesons, and we will not dwell on it any further. 

To substantiate the consideration we pick out ad hoc a few operators of
interest: 
\begin{align}
\nonumber
O_1 & = e m_c \bar{c} \, i\sigma_{\alpha
  \beta} F^{\alpha \beta} \gamma_5 u \,,
& O_2 & = g_s m_c \bar{c} \, i\sigma_{\alpha
  \beta} G^{\alpha \beta} \gamma_5 u \,, \\
O_3 & = [\bar{c}\Gamma_\mu u]([\bar{s}\Gamma^\mu s] +
[\bar{d}\Gamma^\mu d]),
& O_4 & = (\bar{c}\gamma_\mu(1\!+\!\gamma_5)u) \,
(\bar{d}\gamma^\mu(1\!-\!\gamma_5) d)
\label{612}
\end{align}
and put
\beq
{\cal L}_{\rm np} = -\frac{G_F}{\sqrt{2}} \sin{\theta_C}\cos{\theta_C} 
\sum_k c_k O_k,
\label{613}
\eeq
with $c_k$ dimensionless.
The first two operators are the unique quark bilinears.
Operator
$O_3$ has been picked since it evidently represents the CP-odd operator
$o_2$ of the SM -- yet with an arbitrarily inflated coefficient. Consequently,
we would roughly expect 
\beq
|\Im c_3| \approx 10 \cdot \mbox{$\frac{1}{2}$}  r_{\rm SM}\,
\label{614}
\eeq
if $O_3$ is the only New-Physics source of CP violation. 

The operator $O_4$ is an example with a different chiral content 
for both charm and light quarks and differs also in color and
flavor. Operator $O_2$, like $O_3$ is a $U$-spin singlet. 
For the
sake of definiteness we assume in what follows that the 
direct CP asymmetry is largely seen in the $\pi^+\pi^-$ mode having a
numerically smaller SM CP even amplitude, Eq.~(\ref{432}).

The $O_4$ matrix element can be estimated with simple factorization
yielding
\beq
\matel{\pi^+\pi^-}{O_4}{D}\approx 
-i \,f_\pi f_+^{D\tto\pi}(0) M_D^2 \;\frac{1}{N_c}\,\frac{2m_\pi^2}{(m_u\!+\!m_d)m_c}.
\label{860}
\eeq
It would have shown a relative enhancement if charm mass scale were lower,
while would have been suppressed for larger $m_c$. For actual quark 
masses the corresponding factor is not too far from unity. This amplitude is color
suppressed, therefore the factorization is not expected to be a good
approximation -- yet it makes explicit the expected qualitative features required 
for the scale estimates. 

The amplitudes for operators $O_1$ and $O_2$ cannot be estimated by 
simple vacuum insertion. Keeping in mind that both are color-allowed
we use instead a ``rule of thumb''  for our estimates 
\beq
\frac{\matel{\pi^+\pi^-}{O'}{D}}{\matel{\pi^+\pi^-}{O}{D}}\approx
\sqrt{\frac{\Gamma^{\rm part}_{O'}}{\Gamma^{\rm part}_{O}}}
\label{841}
\eeq
and set, as the reference, the operator $O$ to be the `valence' part of $o_1$, 
$(\bar{c}\Gamma_\mu d) \,(\bar{d}\Gamma_\mu u)$ (in fact, its $P$-odd part 
only). In other words, the fraction
of the decay events into the exclusive $\pi\pi$ final state is assumed the
same in the decays mediated by $O$ and $O'$. Since charm mass lies in the
intermediate domain, there must be no large kinematic factors floating
around. 

For $O_2$, the total decay width mediated by $m_c \bar{u}
g_s i \sigma_{\mu\nu}G^{\mu\nu}\gamma_5 c$ is 
\beq
\Gamma_{\sigma G}= 
\alpha_s m_c^5 \frac{N_c^2\!-\!1}{N_c},
\label{840}
\eeq
and the resulting estimate reads
\beq
\matel{\pi^+\pi^-}{m_c\bar{u} g_s i\sigma_{\mu\nu}G^{\mu\nu}\gamma_5 c}{D}\approx
i \,4\pi g_s\sqrt{3}\,f_\pi f_+^{D\tto\pi}(0) M_D^2.
\label{842}
\eeq
(The amplitude proportional to only the first power of $g_s$ 
reflects the fact that we are not yet in the asymptotic heavy quark regime.)

In the case of photonic $O_1$ the partonic rate itself describes the
probability of a different
process, $D\tto \gamma\!+\!\mbox{hadrons}$. Instead we need the similar partonic
rate for the photon conversion into a $d$-quark pair: 
\beq
\Gamma_{\!\rm conv}\!=\! \int \! \frac{{\rm d}\lambda^2}{\lambda^2}\: 
\Gamma_{\sigma F}(\lambda^2)\cdot \frac{\alpha}{3\pi}N_c\,
q_d^2\, , 
\quad  \Gamma_{\sigma F}(\lambda^2) \!=\! 
\frac{e^2m_c^5}{2\pi} 
\,\left(1\!-\!\frac{\lambda^2}{m_c^2}\right)^2
\! \left(1\!+\!\frac{\lambda^2}{2m_c^2}\right)
\theta(m_c^2\!-\!\lambda^2).
\label{850}
\eeq
The kinematic integral equals to 
$\ln{m_c^2/m^2}\!-\!\frac{4}{3}+...$; the lower cutoff can be taken at 
$m$ around $400\MeV$, 
to match the overall hadronic polarization contribution to charge 
renormalization. Then the integral turns out numerically close to unity.
Using this as a counterpart to Eq.~(\ref{840}) we get 
\beq
\matel{\pi^+\pi^-}{m_c\bar{u} e\, i\sigma_{\mu\nu}F^{\mu\nu}\gamma_5 c}{D}\approx
i 8\sqrt{2}\,\pi \alpha \,q_d \,f_\pi f_+^{D\tto\pi}(0) M_D^2 \, ,
\label{852}
\eeq
where the small deviation of the explicit log factor from unity has been
neglected. 

To cross-check the meaningfulness of the estimates we explored alternative
ways. For the gluon bilinear $O_2$ this was relating the gluon field operator
to the chromomagnetic (or kinetic) expectation value in heavy mesons, i.e.\
treating it as fully nonperturbative. This resulted in a different estimate 
\beq
\matel{\pi^+\pi^-}{m_c\bar{u} g_s i\sigma_{\mu\nu}G^{\mu\nu}\gamma_5 c}{D}
\approx
i\frac{2\mu_G^2}{f_\pi}f_+^{D\tto\pi}(0) M_D^2.
\label{832}
\eeq
Assuming numerically that $N_c\alpha_s\!\simeq\! 1$ we would get a number
only $10\%$ smaller than Eq.~(\ref{842}). The two totally
different estimates led to close values because charm lies in between 
the light- and the heavy-quark regimes where both perturbative
partonic and nonperturbative description can qualitatively be applied.

In the case of $O_1$ photon cannot per se be nonperturbative; instead one 
can consider the photon loop with the gluon emitted internally. This effect is
proportional to $q_u$ and is therefore physically different. It describes the
order-$\alpha$ operator mixing of the electromagnetic $O_1$ into chromomagnetic $O_2$. 
Replacing $\ln{\Lambda_{\rm UV}}$ by unity we would obtain 
\beq
\matel{\pi^+\pi^-}{m_c\bar{u} e\, i\sigma_{\mu\nu}F^{\mu\nu}\gamma_5 c}{D}\approx
i 2\sqrt{3}\,q_u\, \alpha g_s \,f_\pi f_+^{D\tto\pi}(0) M_D^2.
\label{847}
\eeq
Comparing this to  Eq.~(\ref{852}) we find the overall factor different: 
\beq
4\sqrt{3\pi \alpha_s} \, q_u\;\mbox{ vs.\ } \;8\sqrt{2}\,\pi q_d,
\label{854}
\eeq
with the former `direct' contribution expectedly dominating since it does not
suffer an additional perturbative loop factor for gluons. This is partially
offset by the larger $u$-quark charge. Altogether the direct photon conversion
estimate appears a few times larger, and we adopt Eq.~(\ref{852}) 
for the $O_1$ estimates. 

Collecting all the expressions we arrive at 
\begin{align}
\nonumber
|\Im c_1| & \approx 
\frac{5.2 \!\cdot\! 10^{-2}}{|\sin{\delta_{\rm FSI}}|}\,, 
& |\Im c_2| & \approx
\frac{0.10 \!\cdot\! 10^{-3}}{|\sin{\delta_{\rm FSI}}|}\,, \\
|\Im c_3| & \approx  
\frac{2 \!\cdot\! 10^{-3}}{|\sin{\delta_{\rm FSI}}|}\,,
& |\Im c_4| & \approx
\frac{4.6 \!\cdot\! 10^{-3}}{|\sin{\delta_{\rm FSI}}|}\,,
\label{862}
\end{align}
assuming that a particular operator is the sole 
source of the New-Physics CP violation.

\section{The Neutron EDM} 

EDMs in general and specifically the neutron EDM $d_n$ are very 
sensitive probes for physics
beyond the SM, in particular for CP violation. As already mentioned, a 
flavor-diagonal CP violation with a size of coupling found for the new physics
operators in the last section would grossly violate the bound, which currently
lies at \cite{gren2006}
\begin{equation}
|d_n| \le 2.9 \!\cdot \! 10^{-26\,} \mathrm{e} \!\cdot\! \mathrm{cm}   . 
\label{864}
\end{equation}
In the following we assume that only the $\Delta C \!=\! \pm 1$ operators
induce the non-SM CP violation, and estimate their
effect.  First we recapitulate the salient points of the 
estimates within the Standard
Model, where recently a new perspective has been proposed \cite{edmsm}.  
We will discard the possibility of strong CP violation assuming
that the long-standing strong CP problem will find a solution where the
QCD $\theta$-parameter is sufficiently close to zero.

\subsection{The Neutron EDM in the Standard Model}
\label{smedm}

The estimates of the neutron EDM in the SM have a thirty year long history; 
the modern perspective can be found in review \cite{pospritz}. EDMs may emerge
from the second order on in the weak interaction and are generally
proportional to $G_F^2$. 

Motivated by the qualitative success of the
constituent quark models in understanding of the properties of 
hadrons, the early estimates of nucleon EDM focused primarily on the EDMs 
of quarks $d_q$. It turned out that for quarks the KM
prediction is further suppressed: the sum of all the two-loop diagrams
vanishes and $d_q$ emerge first at the three-loop
level where an additional loop with at least a gluon must be included
\cite{shabalin}. On top of this, the quark EDM has to be proportional to the
quark mass; this yields an additional suppression for the light quarks.  The
same applies to the color dipole moments of quarks considered as the simplest
induced CP-odd strong force generated through weak interactions at small
distances. The unfortunate feature of the quark EDMs is that there is a
strong numeric cancellation between the leading logarithmic and the
subleading terms in the dominating EDM of the $d$-quark $d_d$ \cite{czar},
which makes it difficult to to make a definite prediction beyond an estimate
$$
|d_{u,d}|\lsim 0.5\cdot 10^{-34\,} \mbox{e}\!\cdot \!\mbox{cm}. 
$$

It has been noted a while ago \cite{gavela,khripzhit} that the strong
suppression intrinsic to $u$ and $d$ quarks can be vitiated in composite
hadronic systems like nucleons.  The transition dipole moments changing
$d$-quark into $s$-quark, electromagnetic or color, are suppressed by the
strange quark mass $m_s$, and such flavor-changing transition without a quark
charge change are already in the loop-induced short-distance renormalization
of the bare weak interaction due to the so-called Penguin diagrams
\cite{penguins}.  

\begin{figure}[hhh]
\vspace*{-.5pt}
 \begin{center}
\includegraphics[width=7cm]{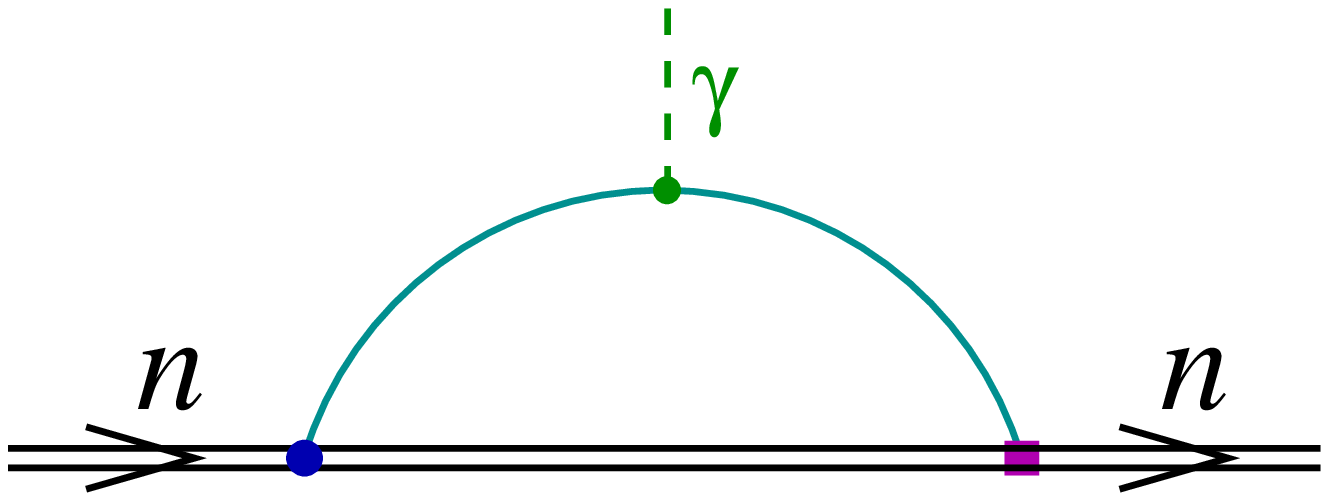}\begin{picture}
(0,0)(0,0)
\put(-37.5,4){\Large ${\bf\Sigma^-}$}
\put(-47,9){\large ${\rm \pi^+}$}
\end{picture}\vspace*{-15pt}
 \end{center}
\caption{ \small
The chirally singular diagram exemplifying the conventional Penguin-based
contribution. One of the vertices is the usual CP-conserving weak amplitude 
while another contains the CP-odd Penguin-mediated operators.}
\label{pisigma}
\end{figure}

It is notoriously difficult to account for the long-range part of the strong
interactions generating the neutron EDM \cite{pospritz}. Usually it is
considered that the principal effect comes from the diagram
in Fig.~\ref{pisigma} having a chiral singularity, or that at least it fairly
represents the magnitude of $d_n$. One of the vertices in the diagram is a
conventional CP-conserving $\Delta S\!=\!1$ weak interaction while the second
is the CP-odd Penguin-induced amplitude originating from short distances
which naturally incorporates heavy quarks, in particular top. 

\thispagestyle{plain}

It has recently been argued \cite{edmsm} that there is a complementary 
mechanism generating $d_n$ to the second order in $G_F$ which does not 
involve short-distance loop effects and is likewise free from chiral 
and $SU(3)$ suppression. It scales like $1/m_c^2$ and would fade out quickly
for sufficiently heavy charm, yet it may actually dominate $d_n$ in the SM 
since charm is marginally heavy in the hadronic mass scale. It originates at
the energy scale around $m_c$ due to interference of the conventional
$\Delta C\!=\!1$, $\Delta S\!=\!0$ weak amplitudes, much in the same wave as
the CP-odd $D$-decay asymmetry discussed in Sect.~\ref{cpvsm}. Consequently,
this mechanism would be present, with a modified strength, in the BSM scenarios 
affecting $\Delta a_{\rm CP}$. The analysis of the BSM contributions to
$d_n$ presented in Sect.~\ref{bsmedm} parallels the SM case, therefore we 
remind below the main steps of Ref.~\cite{edmsm}.

The observable CP-odd effects appear in the second order 
in ${\cal L}_w$ and thus are proportional to $G_F^2$, being embodied in 
\beq
{\cal L}_2= \frac{G_F^2}{2} \!\int {\rm d}^4x \: \mbox{$\frac{1}{2}$}
iT\,\{{\cal L}_w(x)\,{\cal L}_w(0)\}.
\label{108}
\eeq
The generalized GIM-CKM mechanism ensures that the CP-odd piece of 
${\cal L}_2$  is finite in the local four-fermion approximation. The
conventional form of ${\cal L}_w$ applies to the high normalization point around
$M_W$. We will 
neglect, for the most of the consideration, the perturbative gluon
corrections, since the effect exists even without loops. This makes the
analysis simpler and more transparent. 

Descending to a low normalization point we first integrate out top quark and
at the second stage, below $m_b$ also the bottom quark. At tree level
integrating them out simply means discarding all the terms 
containing $t$- or $b$-quark fields. As a result, we arrive at ${\cal L}_2$
generated by a superficially two-family weak Lagrangian
\beq
{\cal L}_w = J_\mu^\dagger J^\mu \mbox{~ with ~}
J_\mu = V_{cs}\, \bar{c} \Gamma_\mu s + V_{cd}\, \bar{c}
\Gamma_\mu d  + 
V_{us}\, \bar{u} \Gamma_\mu s + V_{ud}\, \bar{u}
\Gamma_\mu d \,. 
\label{110}
\eeq
In fact, this is not a true two-family case, since the 
four $V_{kl}$ do not form a unitary matrix; in particular, it is not 
CP-invariant. The phases in the four remaining CKM couplings
cannot all be removed simultaneously by a redefinition of the four quark
fields, as quantified by $\Delta$ in Eq.~(\ref{402}).

\begin{figure}[hhh]
\vspace*{-.5pt}
 \begin{center}
\includegraphics[width=12.cm]{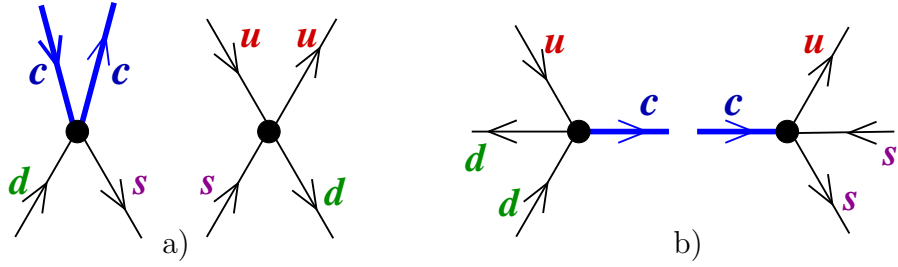}\begin{picture}
(0,0)(0,0)
\put(-99,-1){a)}\put(-31,-1){b)}
\end{picture}\vspace*{-15pt} 
 \end{center}
\caption{ \small
Two types of CP-odd terms. Weak vertices must be off-diagonal in flavor, 
either for down-type (a) or up-type (b) quark. 
Solid dots denote the four-quark vertices. Light lines
correspond to $u$, $d$ or $s$ quarks, thicker lines stand for charm.
 }
\label{lwxlw}
\end{figure}

Interested in $d_n$ we need to consider only the terms in ${\cal L}_2$ that
are diagonal in all four quark flavors. Moreover, we can omit 
explicitly CP-invariant terms of the form of a product of an operator with its
conjugated. Only two operators remain after this selection, those 
proportional to the CKM product $V_{cs}^* V_{us} V_{cd} V_{ud}^*$, and also 
their Hermitian conjugated. 
These non-local $8$-quark operators include both $q$ and $\bar{q}$ fields 
for each of the four quark flavors: 
the CP-odd invariant $\Delta$ (as well as CP-violation
altogether) vanishes wherever any single CKM matrix element becomes zero.
The two terms in ${\cal L}_2$ differ by the type, up- or
down-, of the quark-antiquark pairs coming off the same weak vertex, see Fig.~\ref{lwxlw}. The CP-odd
amplitudes conventionally considered for $d_n$ are of type a) where one of the
weak vertices has $\bar{c}c$ and another  $\bar{u}u$, and both have $\Delta
S\!=\!-\Delta D\!=\!1$. The type b) amplitude has the $\bar{d}d$ and
$\bar{s}s$ pairs in the two weak vertices, respectively, while each 
has $|\Delta C|\!=\!1$. These were routinely omitted. 

Considering the nucleon amplitudes we need to eventually integrate
out the charm quark field as well. At this point the distinction between 
the two types of terms becomes important. Where the two charm fields belong 
to the same four-fermion vertex in the product Eq.~(\ref{108}), Fig.~\ref{lwxlw}a,
they can be contracted into the short-distance loop yielding, for instance,
the usual perturbative Penguins. These are the conventional source of the
long-distance CP-odd effects \cite{gavela,khripzhit}. The loop cannot be formed
for the alternative possibility where $c$ and $\bar c$ belong to different
${\cal L}_w$, Fig.~\ref{lwxlw}b since the charm quark must propagate between
the two vertices. This is the reason why such contributions were usually discarded.

\thispagestyle{plain}

Nevertheless, the latter term has an advantage: it does not involve
short-distance loops, and has a single charm propagator, 
although highly virtual in the hadronic scale. Each weak vertex contains a
flavorless  quark-antiquark pair, but these are light down-type quarks $d$ and
$s$ and are not contracted via a perturbative loop; they will go instead 
into the nucleon wavefunction. The corresponding operator is 
\beq
\frac{G_F^2}{2} \: V_{cs}V_{cd}^* V_{ud}V_{us}^* \int {\rm d}^4x\; 
iT\{(\bar{d}\Gamma_\mu c) (\bar{u}\Gamma_\mu d)_0 \cdot 
(\bar{c}\Gamma_\nu s)  (\bar{s}\Gamma_\nu u)_x \, +\, \mbox{H.c.}
\label{124}
\eeq
The Hermitian conjugate, apart from complex conjugation of the CKM product, is
simply the exchange between $s$ and $d$, $s\!\leftrightarrow\! d$ (this
particular property does not hold beyond the SM). For the
sake of transparency, we have passed here to the sum
and the difference of the two operators in ${\cal L}_w$, in terms of 
Eq.~(\ref{404}).

As the space separation $x$ in Eq.~(\ref{124}) is of order $1/m_c$,
eliminating charm results in a local OPE; the expansion parameter $\mhad/m_c$
is not too small and we need to keep a few first terms. The tree-level OPE is
particularly simple here and amounts to the series
\beq
c(0)\bar{c}(x)= \left(\frac{1}{m_c\!-\!i\!\not\!\!D}\right)_{0x}= 
\frac{1}{m_c} \delta^4(x) + \frac{1}{m_c^2}\delta^4(x)\, i\slashed{D} + 
\frac{1}{m_c^3}\delta^4(x)\, (i\slashed{D})^2 +...
\label{128}
\eeq
valid under the $T$-product. For purely left-handed weak currents in the SM
the odd powers of $1/m_c$ in Eq.~(\ref{128}) are projected out, including the
leading $1/m_c$ piece.  We then retain only the $1/m_c^2$ term and arrive at the
local effective CP-odd Lagrangian\footnote{We have changed notations compared
  to Ref.~\cite{edmsm}: now $\tilde O_{uds}$, $O_{uds}$ and $O^\alpha_{uds}$
  all include subtraction of the Hermitian conjugated operator.}
\bea
\label{130}
\tilde{\cal L}_- \msp{-4}&=&\msp{-4} -i\Delta  
\frac{G_F^2}{2m_c^2}\,\tilde O_{uds} , \\
\nonumber
\tilde O_{uds} \msp{-4}&=&\msp{-4} 
(\bar{u}\Gamma^\mu d)\, (\bar{d} \Gamma_\mu i\slashed{D}
\Gamma_\nu s)\,(\bar{s}\Gamma^\nu u) - \{d\leftrightarrow s\} \!=\! 
\\
\nonumber
\msp{-4}& &\msp{-4}
(\bar{u}\Gamma^\mu d)\!\cdot\! 
\left[(\bar{d} \Gamma_\mu i\slashed{D}
\Gamma_\nu s)\!\cdot\!(\bar{s}\Gamma^\nu u) \!+\!  
(\bar{d} \Gamma_\mu i\gamma_\alpha
\Gamma_\nu s)\,i\partial^\alpha(\bar{s}\Gamma^\nu u)
\right]
- \{d\leftrightarrow s\}
;
\eea
in the last expression the covariant derivative acts only on the
$s$-quark field immediately following it. 

To address the electric dipole moments we need to incorporate the electromagnetic
interaction. One photon source lies in the covariant derivative in the
operator $\tilde O_{uds}$, which includes electromagnetic potential along with
the gluon gauge field. It is proportional to the up-type quark electric charge
$+\frac{2}{3}$. The corresponding photon vertex is local and is given by the
Lorentz-vector six-quark operator which we denote as $O^\alpha_{uds}$:
\beq
O^\alpha_{uds}\!=\! (\bar{u}\gamma^\mu (1\!-\!\gamma_5)s)\, 
(\bar{s} \gamma_\mu i\gamma^\alpha
\gamma_\nu (1\!-\!\gamma_5) d)\,(\bar{d}\gamma^\nu (1\!-\!\gamma_5) u)- 
\{d\leftrightarrow s\} .
\label{650}
\eeq
Another, non-local contribution is the $T$-product of the pure
QCD part of $\tilde O_{uds}$ 
\beq
O_{uds}\!=\! (\bar{u}\gamma^\mu (1\!-\!\gamma_5) s)\, 
(\bar{s} \gamma_\mu i\slashed{D}
\gamma_\nu (1\!-\!\gamma_5) d)\,(\bar{d}\gamma^\nu (1\!-\!\gamma_5) u) - \{d\leftrightarrow s\}
\label{651}
\eeq
with the light-quark electromagnetic 
current.\footnote{In general only the sum of the two terms yields the
transverse electromagnetic vertex; however, when projected on the dipole moment
Lorentz structures they separately conserve current.}
The total photon vertex is thus given by the effective CP-odd 
Lagrangian
\beq
 A_\alpha{\cal L}_-^\alpha \!=\! -e \,i\Delta \frac{G_F^2}{m_c^2} A_\alpha \!\left[
\mbox{$\frac{2}{3}$} O_{uds}^\alpha \!+ \!\int \!\!{\rm d}^4 x \:iT\{O_{uds}(0)\; 
J^\alpha_{\rm em}(x) \}\right]\!, 
\quad\: J^\mu_{\rm em}\!=\!\sum_q \mbox{$\frac{e_q}{e}$} \,\bar{q}\gamma^\mu q, 
\label{132}
\eeq 
where $A_\mu$ is the electromagnetic potential and $e$ is the unit charge. 

In principle, the local and non-local pieces above correspond to distinct
physics: one has photon emitted from distances of order $1/m_c$ while the
latter senses long-distance charge distribution.  The latter
usually dominates, however the specifics of the left-handed weak interactions
in the SM make them of the same $1/m_c^2$ order.

An important feature of the considered contribution is that it remains
finite in the chiral limit and it does not vanish if $d$ and $s$ quarks become
nearly degenerate, at first glance contradicting 
the origin of the KM mechanism where an additional $SU(2)$ freedom to mix $s$
and $d$ makes the theory CP-invariant at
$m_s\!=\!m_d$.\footnote{The contribution would vanish if charm
and top become degenerate; considering the cases of degenerate bottom and
strange quarks, or charm and up makes no sense in this context since it has
been assumed as the starting point that $m_b, m_c \gg \mhad$ while $u$, $d$ and
$s$ are light quarks.} This in fact is fully consistent, since the external
state, the neutron, is explicitly $s\leftrightarrow d$ non-symmetric, and 
would not stay invariant under the mixing transformation. The same applies,
for instance, to the $d$-quark EDM.
In contrast, short-distance effects of light quarks in the loops 
involve severe GIM-type suppression proportional to the powers of the light 
quark masses. The quark electric dipole moments as the purely short-distance
contributions are explicitly proportional to the corresponding quark mass
for chirality reasons. 

Let us parenthetically note that there is no a formal contradiction between the
nonvanishing expression for $d_n$ and the explicit T-invariance at literally 
$m_s\!=\!m_d$ either. $T$-invariance
prohibits dipole moments only for the eigenstates of the
Hamiltonian; for instance, non-diagonal dipole moment matrix elements 
are perfectly allowed. 
The above
considered neutron, a baryon state with strangeness $S\!=\!0$ is such a
physical eigenstate only as long as the mass splitting $m_s\!-\!m_d$ remains
much larger than the weak corrections to the hadron masses, since weak
interactions violate flavor. Where  $m_s\!-\!m_d$ becomes of the order of
$G_F m_q m_c^2\,$ mixing between neutron and its strange partners must be
accounted for. In this regime the time-violating EDM for a physical state
should be distinguished from the conventional $d_n$; the former would vanish at 
$m_s\!=\!m_d$.

The CP-odd operators contain strange quark fields. This means that the induced
effects would vanish in a valence approximation to nucleon where only $d$ and
$u$ quarks are active.  It is known, however, that even at low normalization
point the strange sea in nucleon is only moderately suppressed. The
large-$N_c$ perspective on the nucleons paralleling the picture of the baryon
as a quantized soliton of the pseudogoldstone meson field \cite{soliton} makes
this explicit: the weight of the operators with strange quarks in the chiral
limit is generally determined simply by the operator-specific Clebsh-Gordan
coefficients of the $SU(3)$ group.  

Such an `intrinsic  strangeness' suppression is specific for the considered
mechanism to generate $d_n$ in the SM; the
conventional contribution trades it in for the `intrinsic charm'.
Associating the virtual-pair suppression with the
strangeness sea in the nucleon is probably a relatively light price to pay.  In
contrast, the perturbative Penguin
effects yield small coefficients whenever considered in the truly
short-distance regime.

\subsubsection{Matrix elements}

The CP-odd operators $O^\alpha _{uds}$ and 
$O_{uds}$ have high dimension; this is routinely associated with 
being poorly defined for practical applications. However,
these particular operators possess intrinsic symmetry properties, including
antisymmetry in respect to $s\leftrightarrow d$, which prohibit mixing with
lower-dimension operators, and make them a suitable object for the
full-fledged nonperturbative analysis.

The neutron EDM is obtained by evaluating the hadronic operators 
in Eq.~(\ref{132}) over the neutron
state. Since ${\cal L}_-^\mu$ is T-odd, the matrix element vanishes for
zero momentum transfer and the linear in $q$ term 
describes $d_n$: 
\beq
\matel{n(p\!+\!q)}{{\cal L}_-^\mu}{n(p)}= d_n\:
q_{\nu\,} \bar{u}_n(p\!+\!q)i\sigma^{\mu\nu}\gamma_5 u_n(p).
\label{654}
\eeq
Neither of the two matrix elements involved are easy to evaluate, although 
one may hope that such a contribution may eventually be determined
without major ambiguity, including the definitive prediction for the overall
sign. Although only the $P$-violating part of  $O^{(\mu)}_{uds}$
contributes, the original form is more compact and makes symmetry explicit.

The contact operator $O^{\mu}_{uds}$ is a product of three left-handed
flavor currents; $O_{uds}$ instead of the $\bar{s}d$ current has a flavor
non-diagonal left-handed partner of the quark energy-momentum tensor in the
chiral limit. Therefore it seems plausible that the required matrix elements
can be directly calculated within the frameworks like the Skyrme model
\cite{skyrme,soliton}, or in its dynamic QCD counterpart \cite{liquid} derived
in the large-$N_c$ limit from the instanton liquid approximation.

Lacking presently more substantiated calculations we resort to the simple
dimensional estimates. 
For the local piece we put
\beq
\matel{n(p\!+\!q)}
{O_{uds}^\mu}
{n(p)} 
\!=\! 2i{\cal K}_{uds}\: q_\nu \bar{u}(p\!+\!q) i\sigma^{\mu\nu}\!\gamma_5 u(p).
\label{660}
\eeq
The reduced matrix element ${\cal K}_{uds}$ has dimension of mass to the fifth 
power. We estimate it as
\beq
|{\cal K}_{uds}| \approx \kappa \;\mhad^5,
\label{662}
\eeq
where $\mhad$ is a typical hadronic momentum scale and $\kappa$ stands 
for the `strangeness suppression' to account for the fact that 
neutron has no valence strange quarks; $\kappa \!\approx \!1/3$ is taken as a
typical guess. 

Due to the high dimension of the operators the 
estimate for $d_n$ depends dramatically 
on the value used for $\mhad$. Although the typical 
momentum of quarks in nucleon is around $600\MeV$ or higher, using this as
$\mhad$ would strongly overestimate the effect. 
Six  powers of mass in Eq.~(\ref{662})
come from the product of two local light quark currents each intrinsically
containing factors $N_c/8\pi^2$ when converted into the conventional momentum
representation. This is illustrated by the magnitude of the vacuum quark
condensate where such a factor effectively reduces $\mhad^3$ down to $\sim\!
(250\MeV)^3$. 

To account for such differences we assign a factor of
$(0.25\GeV)^3\!\equiv\!\mu_\psi^3 $ to each additional quark current in the
product, while the remaining dimension will be made of the powers of $\mhad$
taken around $500\MeV$.
Then this contribution to $d_n$ becomes
\beq
|d_n| = \frac{32}{3} e \,\Delta \,\frac{G_F^2}{m_c^2}  |{\cal K}_{uds}|
\approx  3.3 \cdot 10^{-31} e\!\cdot \!\mbox{cm} 
\times \kappa \left(\frac{\mu_\psi}{0.25\GeV}\right)^6
\left(\frac{0.5\GeV}{\mhad}\right) \!,
\label{664}
\eeq
where $\Delta\!\simeq\!3.4\!\cdot\! 10^{-5}$ has been used. 
An independent enhancement factor may come from summation over the Lorentz indices 
in the currents, but we neglect it.

The most naive estimates for the $d_n$ induced by the non-local
piece in Eq.~(\ref{132}) would yield a similar dimensional scaling 
except that no explicit
factor $e_c\!=\!\frac{2}{3}$ appears: the dimension of the non-local
$T$-product is the same as of $O_{uds}^\mu$ itself. Following the more careful
way advocated above where we distinguish the mass scale associated with 
the local product of the quark fields, the result is literally different:
\beq
|d_n|^{\rm n-loc} \approx e \,\Delta \,\frac{G_F^2}{m_c^2}  
32 \kappa\, \mu_\psi^9 \,\mhad^{-4}
\approx  1.2 \cdot 10^{-31} e\!\cdot \!\mbox{cm} 
\times \kappa \left(\frac{\mu_\psi}{0.25\GeV}\right)^9
\left(\frac{0.5\GeV}{\mhad}\right)^4\!;
\label{665}
\eeq
numerically the difference is not radical, however.

Alternatively, the non-local contributions can be analyzed focusing on
the contributions of the individual intermediate states, usually the
lowest in mass.  Among them the
hidden-strangeness states, including $\bar{K}
\Lambda(\Sigma)$ look promising suggesting a way to dynamically 
estimate the `intrinsic strangeness'  factor $\kappa$. In the standard model,
however, the corresponding loops are not infrared-enhanced and rather saturate
at large virtual mass yielding a result strongly dependent on the assumed
cutoff. For the same reason the kaon and the lowest baryon as the intermediate
state are not any more remarkable a priori than ordinary resonances.  

\begin{figure}[hhh]
\vspace*{-5pt}
 \begin{center}
\includegraphics[width=10.cm]{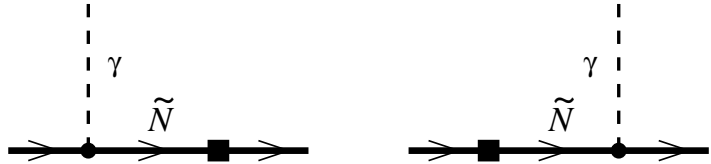}\vspace*{-18pt}
 \end{center}
\caption{ \small
Nonlocal contribution to $d_n$ with the intermediate $\tilde N$. Solid block
denotes the CP-odd operator $O_{uds}$. 
}
\label{nonloc}
\end{figure}

Ref.~\cite{edmsm} considered the contribution of the lowest resonant 
state, the $\frac{1}{2}^-$ nucleon resonance $N(1535)$ referred to below 
as $\tilde N$, as an alternative estimate of the non-local piece in $d_n$.
In terms of the two hadronic vertices, 
\beq
\matel{n(p')}{J_{\rm em}^\mu (0)}{\tilde N (p)}\!=\! -\rho_{\tilde N}
\bar{u}_n i\sigma^{\mu \nu} \gamma_5 q_\nu  u_{\tilde{N}} , \quad 
\matel{\tilde N(p')}{O_{uds} (0)}{n(p)}\!=\! 16 i
{\cal N}_{\!uds}\bar{u}_{\tilde{N}} u_n
\label{666}
\eeq
the sum of the Feynman diagrams in Fig.~\ref{nonloc} gives
\beq
d_n^{(\tilde N)} = -e \: \Delta \frac{32 G_F^2}{m_c^2} 
\left(\frac{  \rho_{\tilde N}\, {\cal N}_{\!uds}}
{M_{\tilde{N}}\!-\! M_N} \right) .
\label{668}
\eeq
The electromagnetic vertex estimated from the measured transition 
$ \tilde{N} \tto n\!+\!\gamma$ becomes $\rho_{\tilde N} \!\approx\! (0.34 \pm
0.08) \GeV^{-1}$. The induced weak CP-odd vertex is estimated in the
dimensional way, for the dimension-ten operator $O_{uds}$ yielding
\beq
|{\cal N}_{\!uds}| \approx \kappa \;\mu_\psi^6 \,\mhad.
\label{670}
\eeq
Finally this estimate reads
\beq
|d_{n}|^{(\tilde N)} \!\approx \! e \, \Delta \frac{32 G_F^2}{m_c^2}  
\kappa\,\mu_\psi^6 \,\mhad \, \frac{\rho_{\tilde N} }{M_{\tilde{N}}\!-\! M_n} 
\approx \!
1.4 \cdot 
10^{-31} \mbox{e$\cdot$cm} \times \kappa \left( \frac{\mu_\psi}{\mbox{0.25\GeV}}
\right)^6 \!\left( \!\frac{\mhad}{\mbox{0.5\GeV}} \!\right)
. 
\label{674}
\eeq
This value is consistent with the direct dimensional estimate of
the non-local contribution, in particular considering 
the fact that the lowest excited state alone may not necessary saturate it. 
Therefore, in further applications we generally follow the more 
straightforward estimates paralleling  Eq.~(\ref{665}). 

\thispagestyle{plain}

Finally, our estimate for $d_n$ in the SM centers around 
$10^{-31\,}\mbox{e$\cdot$cm}$ although even the values $5$ to $10$ times larger
may not be excluded.

The natural benchmark for the CKM $d_n$ in the SM evidently lies about
\beq
d_n \propto \Delta\, G_F^2 \mhad^3.
\label{682}
\eeq
In the same terms the loop-less contribution considered above is
\beq
d_n \propto \Delta\, G_F^2 \mhad^3 \!\cdot\! \frac{\mhad^2}{m_c^2} \!\cdot\! \kappa.
\label{683}
\eeq
The last factor reflects the absence of valence strange quarks in the nucleon.
The related suppression is unavoidable for $d_n$ in one form or another; it is 
natural to
think that paying the price by the soft strangeness content in the nucleon
state is the minimal burden. Therefore, from this perspective such a
contribution appears to bear a mild model-specific suppression, since
$m_c$ in practice only moderately exceeds the characteristic hadronic scale
$\mhad$. 

At the same time $d_n$ does not contain parametric chiral enhancement,
$\ln{\mhad^2/m_\pi^2}$ or numerically significant scalar matrix
elements possible in the case of generic couplings. This implies also a loss
of a potential factor of a few; it may be recovered by the interactions
generated beyond the SM.

Throughout the long history of the conventional long-distance contributions to
$d_n$ in the SM it has usually been considered  \cite{pospritz} that the 
principal effect comes from the diagram in Fig.~\ref{pisigma} peculiar by showing a 
chiral singularity. 
It has been calculated in Ref.~\cite{khripzhit},
\beq
d_n \approx e G_F^2 \Delta \frac{C_{\rm pert}\alpha_s}{27\sqrt{2}
  \pi^3}\ln{\frac{m_t^2}{m_c^2}} \frac{2|\!\aver{\bar\psi \psi}\!|\, m_\pi^2}{f_\pi m_s}
\tilde A (2\alpha\!-\!1)g_A \ln{\frac{m_K}{m_\pi}}
\label{692}
\eeq
($\tilde A$ is a strong constant parameterizing the conventional CP-even
vertex and $\alpha$ a dimensionless ratio of two $SU(3)$ meson-to-baryon
axial couplings, while $C_{\rm pert}$ stands for additional perturbative
factors). The original authors' estimate was close to $d_n\!\approx \!
2\!\cdot\! 10^{-32} \mbox{e$\cdot$cm}$. It was done in 1981 when even the
size of the CKM admixture of the third generation was unknown and was 
thought to be of the scale of $\theta_C$. The equivalent of the CP-violating 
parameter $\Delta$ likewise was estimated assuming $m_t\!\approx\! 30\GeV$, yet
a value only $1.5$ times larger than known today, see 
Eq.~(\ref{402}), was used. At the same time the used log ratio
of the $t$ and $c$ quark masses was somewhat smaller. 
The size of this contribution is now usually cited as 
$d_n\!\approx \!10^{-32\,} \mbox{e$\cdot$cm}$ \cite{pospritz}.

The $d_n$ value Eq.~(\ref{692}) is  proportional to $\alpha_s/\pi$ 
from the short-distance Penguin loop. It also contains a
factor $m_\pi^2 \!\propto\! m_q$ compared to the benchmark
Eq.~(\ref{682}), however the overall light-quark mass scaling is remarkable:
$m_{u,d}$ enter divided by $m_s$ rather than by $\mhad$. Therefore, in the
$SU(3)$ chiral limit where all $m_{u,d,s}\tto 0$, $m_q/m_s$ fixed it would stay
finite. 

In practice, however the $SU(2)$ chiral limit  $m_{u,d}\tto 0$, $m_s$ fixed is
more relevant in numeric estimates. In this case Eq.~(\ref{692}) has an
additional factor of the light-quark mass $m_q$ compared to the benchmark
value Eq.~(\ref{682}). This is in agreement with the general fact stated in 
Appendix~\ref{ggt}: the contribution contains a chiral log and 
therefore must include an explicit
factor of $m_q$ since the SM weak amplitudes do not contain right-handed light
quark fields. As emphasized there, it is sufficient to check this for
the bare weak vertices. 

This illustrates the underlying problem in estimating $d_n$ in the SM: the
physically distinct chirally singular contributions have to be
$m_q$-suppressed. The leading-$m_q$ contributions are not related to soft
pions and are rather saturated at the loop momenta of the typical hadronic
mass scale $\mhad$, or by resonances with a significant mass gap. Such effects
are generally uncertain and may involve cancellations.

The conventional SM contribution  Eq.~(\ref{692}), therefore, has an
additional light-mass suppression $\propto \!m_q$ on top of the perturbative
short-distance factor. Although it is partially offset by numerically large
factors accompanying the amplitudes with right-handed light quarks, together
with the perturbative loop factor it results in a certain suppression. This may
explain the larger number for the loopless EDM which we estimate to be around
$10^{-31} \mbox{e$\cdot$cm}$.

\subsection{Neutron EDM and a BSM Charm CP Violation}
\label{bsmedm}
In order to estimate the effect of the $|\Delta C| \!=\! 1$ amplitudes on the
neutron EDM we should replace one of the two ${\cal L}_w$ in the
product ${\cal L}_2$ 
by the New Physics operators. We will assume that we
get a reasonable estimate when considering one operator at a time; this basically 
corresponds to the assumption that in the neutron EDM we do not have
a destructive interference absent from the charm decays.

As is clear from the analysis of the SM case, the operator structure obtained
upon integrating out charm depends on its chirality in the NP amplitude: 
in the left-handed case it follows the SM case. Where the charm field is
right-handed, only the odd-power terms $1/m_c$, $1/m_c^3$... survive. In this
case the leading term suffers less from the $\mhad/m_c$ suppression, however
it does not include the leading contact photon vertex (the photon operator
$O_1$ is an exception in this respect) which appeared to yield a few times larger
contribution, at least within our estimates. The contact photon vertex is then
delayed till order $1/m_c^3$. Such a peculiarity introduces certain difference,
but in view of the relatively mild numeric power suppression the presence of
the nonlocal $T$-product term to the leading $1/m_c$ order for the
right-handed charm does not appear to bring in a notable numeric difference. 

The chiral content of the light valence quarks generally makes a bigger
difference. The nucleon matrix elements with both left-handed and right-handed
fields are usually numerically enhanced as seen on the example of the nucleon
$\sigma$-term. Moreover, a CP-odd scalar pion-to-nucleon coupling 
may be induced. Although the CP-nonconservation case is more
involved, it can be stated that this vertex at small momentum transfer 
would be proportional to the light quark masses and therefore negligible 
in practice unless the New Physics operators include right-handed light 
quarks, see Appendix~\ref{ggt}. 

\begin{figure}[hhh]
\vspace*{-.5pt}
 \begin{center}
\includegraphics[width=5cm]{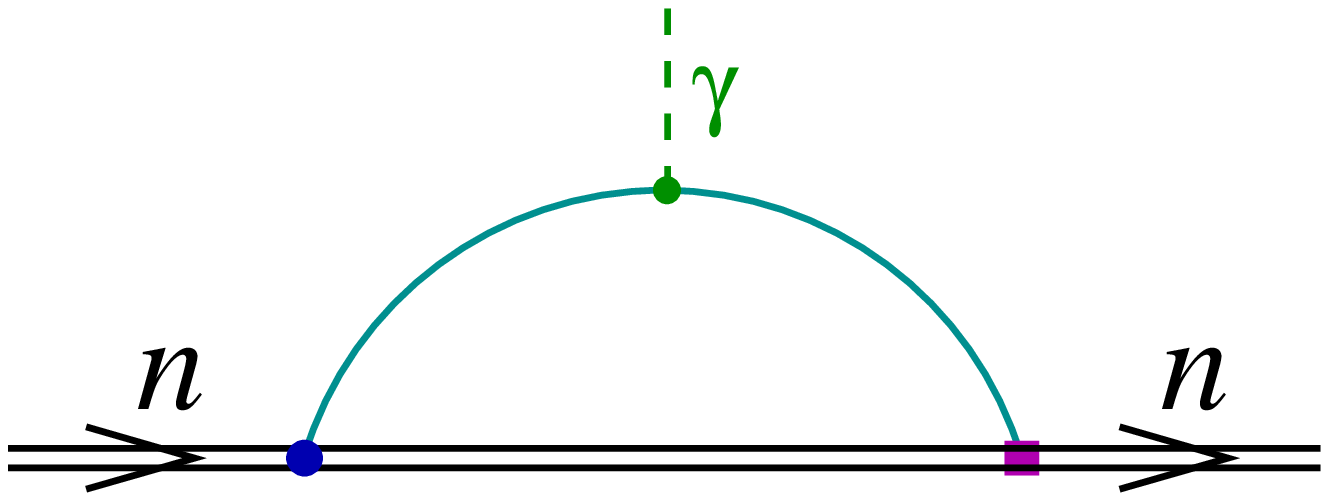}\begin{picture}
(0,0)(0,0)
\put(-26,3.5){\scalebox{1.2}{$p$}}\put(-32,7){\scalebox{.8}{${\rm\pi^-}$}}
\end{picture}\vspace*{-15pt}
 \end{center}
\caption{ \small
The chirally singular diagram generated by the scalar vertex. One of the 
vertices is the usual strong 
CP-conserving pseudoscalar coupling while another is the 
induced CP-odd scalar vertex.  
}
\label{chiloop}
\end{figure}

The contact photon vertex contribution to $d_n$ does not depend on
the induced pion-nucleon interaction. The scalar pion vertex 
${\cal G}_s\,\bar{p} n\pi^-$, on the other
hand,
generates a chirally enhanced long-distance contribution to the $T$-product
piece with the $\pi^-p$ intermediate state, described by the diagrams in
Fig.~\ref{chiloop}: 
\beq
-e\frac{{\cal G}_s g_{\pi NN}}{16\pi^2 M_N}\ln{\frac{\Lambda^2}{m_\pi^2}}\,
\bar u_n i\sigma_{\mu\nu}q^\nu \gamma_5  u_n 
= -e\frac{{\cal G}_s g_A}{8\pi^2 f_\pi}\ln{\frac{\Lambda^2}{m_\pi^2}}\,
\bar u_n \sigma_{\mu\nu}q^\nu i\gamma_5  u_n 
\label{944a}
\eeq
with $\Lambda$ the ultraviolet cutoff.
In the actual world the chiral $\ln{\frac{\mhad^2}{m_\pi^2}}$ constitutes a
moderate factor about $3$ and therefore is remarkable more in the
conceptual aspect. However it comes proportional to the large couplings and this
makes up for the loop factor. In the few  considered examples the overall chiral
enhancement roughly 
offsets the typical suppression (at the same order in $1/m_c$) 
of the $T$-product piece relative to the contact photon vertex contribution.
Regardless of the details, it is clear that the chiral log per se is a too weak
singularity to change dramatically the expected magnitude of $d_n$.

\thispagestyle{plain}

The effect of the scalar pion-to-nucleon vertex can be more pronounced in atomic
EDMs. The pion-mediated nuclear forces are relatively long-range and may be
additionally amplified for the isoscalar coupling in heavy atoms.

The current algebra technique allows to unambiguously determine the induced scalar
vertex in the chiral limit, see Appendix~\ref{ggt}. Of course, in the general
case it would require the nucleon expectation values of similar
high-dimension effective operators, currently estimated in a rather crude
way. An additional uncertainty would come from possible cancellations due to
the second, pole-subtraction term in Eq.~(\ref{2030}). As a rule, aiming only at the
overall magnitude, we simply neglect this term. 

Contracting charm propagator we end up with the multi-quark CP-odd
operators. The neutron EDM induced by them is evaluated applying the dimensional 
estimates elaborated for the SM case; they are described in
Sect.~\ref{smedm}. The non-valence strange quarks are neglected here and no
factor $\kappa$ appears. 

Considering the quark bilinears, we start with the more natural gluonic $O_2$.
In view of the above mentioned difference in the OPE, we consider separately
the cases of $O_2^R$ and $O_2^L$ containing right- and left-handed $c$-field,
respectively, rather than its scalar and pseudoscalar versions.  \vspace*{2pt}

We start with 
$O_2^R\!=\!m_c \bar c g_s i\sigma G (1\!-\!\gamma_5) u$. 
Here charm induces already the $1/m_c$ effect generating $d_n$ via the
$T$-product with the electromagnetic current. The contact term emerges 
only to order $1/m_c^3$, and we discard it. The corresponding CP-odd operator
is 
\beq
O^{(-)}\!=\!-i\,\Im c_2\,\frac{G_F^2\sin^2{\!\theta_c}\cos^2{\!\theta_c}}{m_c^0}\,
\left[\bar{u}\, g_s i\sigma_{\alpha\beta}G^{\alpha\beta}\gamma_\mu(1\!-\!\gamma_5)d
\,\bar{d}\gamma^\mu(1\!-\!\gamma_5)u - d \leftrightarrow s\right] +\mbox{H.c.}
\label{910}
\eeq
For the scale estimate we simply factor out the operator 
$g_s i\sigma_{\alpha\beta}G^{\alpha\beta}$ and assume it has the value similar to 
the one in heavy mesons or baryons, $2\mu_G^2\!\approx \!0.7\GeV^2$. A close
value is obtained if we use the vacuum condensate $\aver{\bar{q}g_s i\sigma G q}
\!\approx \! 0.8\GeV^2 \aver{\bar{q}q}$.
Discarding strange quarks and applying to the rest our 
dimensional estimate we get 
\beq
|d_n|  
\approx  \Im c_2\, G_F^2\sin^2{\!\theta_c}\cos^2{\!\theta_c}\,
 32\mu_G^ 2 \,\frac{(0.25\GeV)^6}{(\mhad)^5}\,\chi_{\rm fl} \approx 
1.1\cdot 10^{-26} e\!\cdot \!\mbox{cm} \cdot 
\Im c_2 \:\chi_{\rm fl}\,,
\label{920}
\eeq
where $\chi_{\rm fl}\!\approx\!1 \,\mbox{to}\,2 $ is a flavor factor which 
accounts for the fact that 
there are two $d$ quarks in the neutron.   
Using Eqs.~(\ref{862}) for $\Im c_2$ we get 
\beq
|d_n| \approx  10^{-30}  \frac{\chi_{\rm fl}}{|\sin{\delta_{\pi^+\pi^-}}|}
  e\!\cdot \!\mbox{cm}\approx   
2.3\chi_{\rm fl}\cdot 10^{-30}e\!\cdot \!\mbox{cm}\: .  
\label{922}
\eeq
Here and in what follows we assume $|\sin{\delta_{\rm FSI}}|\approx 0.5$ as a 
typical value. Thus we expect an enhancement of roughly 
a factor of thirty. 
\vspace*{2pt}

Now we turn to $O_2^L\!=\!m_c \bar c g_s i\sigma G (1\!+\!\gamma_5) u$.
Here the leading term $1/m_c$ vanishes like in the SM and the $1/m_c$
expansion starts with $1/m_c^2$, yet we have the right-handed $u$-quark which
entails chiral enhancements in the nucleon matrix elements. There are 
two distinct contributions to the leading order like in the SM, the contact 
and non-local.  For the latter we obtain 
\bea
\label{912}
O^{(-)}\msp{-4} &=& \msp{-4} 
-i\,\Im c_2\,\frac{2G_F^2\sin^2{\!\theta_c}\cos^2{\!\theta_c}}{m_c^1}\,\times \\
\nonumber
&& \msp{-18.5}
\left[\bar{u}g_s\!\left(\!G^{\mu\nu}\!\!-\!i\tilde G^{\mu\nu}\! 
+\! G^{\mu\alpha}\sigma_{\alpha\nu}
\!-\! \sigma^{\mu\alpha}G_{\alpha\nu}\!+\! \mbox{$\frac{1}{2}$}
\delta^{\mu\nu}\sigma_{\alpha\beta}G^{\alpha\beta} \!\right) 
\!D_\nu (1\!-\!\gamma_5) d
\,\bar{d}\gamma_\mu(1\!-\!\gamma_5)u - d \leftrightarrow s\right]\!+\!\mbox{H.c.}
\eea
and the contact electromagnetic vertex is given by 
\bea
\label{913}
O^{\nu(-)}\msp{-4} &=& \msp{-4} 
-i\,\Im c_2\,\frac{2G_F^2\sin^2{\!\theta_c}\cos^2{\!\theta_c}}{m_c^1}\,\times \\
\nonumber
&& \msp{-18}
\left[\bar{u}g_s\!\left(\!G^{\mu\nu}\!-\!i\tilde G^{\mu\nu}\! 
+\! G^{\mu\alpha}\sigma_{\alpha\nu}
\!-\! \sigma^{\mu\alpha}G_{\alpha\nu}\!+\! \mbox{$\frac{1}{2}$}
\delta^{\mu\nu}\sigma_{\alpha\beta}G^{\alpha\beta}
\right)\! (1\!-\!\gamma_5) d
\,\bar{d}\gamma_\mu(1\!-\!\gamma_5)u - d \leftrightarrow s\right]\!+\!\mbox{H.c.}
\eea
It likewise may have a numeric chiral enhancement due to right-handed
$u$-quark, yet no literal chiral log from the pion loop. Our estimate for it
reads as 
\beq
|d_n|^{\rm loc} \! 
\approx \! \frac{2}{3}\Im c_2\, G_F^2\sin^2{\!\theta_c}\cos^2{\!\theta_c}\,
 \frac{32\mu_G^2}{m_c} \,\frac{(0.25\GeV)^3}{\mhad}\,\chi_{\rm fl}\,
\chi_{\rm scal} \!\approx \!
2.5 \!\cdot\! 10^{-26\,} e \cdot \!\mbox{cm} \cdot 
\Im c_2 \:\chi_{\rm fl}\,\chi_{\rm scal},
\label{924}
\eeq
where we, as above, have  equated the whole bracket containing 
the gluon field strength, including $g_s$, with $2\mu_G^2$; yet another 
factor $\chi_{\rm scal}$ has been added to indicate a possible
enhancement of the scalar expectation value (cf.\ the size of the
nucleon $\sigma$-term).
This is about  $3$ times larger than in Eq.~(\ref{920}). 

The non-local contribution estimated dimensionally is typically $2.5$ to $3$ times
smaller than the contact one. However here the right-handed $u$-quark induces
the nonvanishing scalar pion-nucleon vertex and 
the $\pi^- p$ intermediate state yields a chiral log, cf.\
Eq.~(\ref{944a}). Combined with the current algebra result for the scalar
version as described above this enhancement numerically turns out 
about $3.5$, i.e.\ we  get a number close to the contact estimate Eq.~(\ref{924}).

Thus, we can use for this case the local estimate Eq.~(\ref{924}) and 
$\Im c_2$  from Eq.~(\ref{862}). Then 
\beq 
|d_n| \approx 2.5\!\cdot\! 10^{-30} 
\frac{\chi_{\rm fl}\,\chi_{\rm scal}}{|\sin{\delta_{\pi^+\pi^-}}|} \,
e\!\cdot \!\mbox{cm}\approx \chi_{\rm fl} \,\chi_{\rm scal}
\cdot 5 \!\cdot\! 10^{-30\,}e\!\cdot \!\mbox{cm}\: .
\label{926}
\eeq
For this chiral structure we get about an $80$-fold enhancement compared to the SM. 
\vspace*{3pt}

The photonic operators $O_1$ are the simplest case since only the contact
photon vertex should be considered to the leading order in $\alpha$. 
For the case of the right-handed $c$ quark, 
$O_1^R\!=\!e m_c \bar c i\sigma F (1\!-\!\gamma_5) u$  
the leading term in the $1/m_c$ expansion yields
\beq
O^{\nu(-)}= 2 i\,\Im c_1\,G_F^2\sin^2{\!\theta_c}\cos^2{\!\theta_c}\,
\partial_\mu \left[\bar{u}i\sigma^{\mu\nu} \gamma_\alpha(1\!-\!\gamma_5)d
\,\bar{d}\gamma^\alpha(1\!-\!\gamma_5)u - d \leftrightarrow s\right] +\mbox{H.c.}
\label{930}
\eeq
The matrix elements of the CP-even partner of such an operator may have
been estimated in the literature. Applying our standard recipe we get 
\bea
|d_n| \msp{-7}&&  
\approx  16|\Im c_1|\, G_F^2\sin^2{\!\theta_c}\cos^2{\!\theta_c}\,
(0.25\GeV)^3 \,\chi_{\rm fl}  \nonumber \\ 
&& \approx 
3.2\cdot 10^{-26\,}  e\!\cdot \!\mbox{cm} \cdot  \Im c_1 \!\cdot\!\chi_{\rm fl}
\approx 3.4\!\cdot\! 10^{-27}  e\!\cdot \!\mbox{cm}\!\cdot\!\chi_{\rm fl}\,.
\label{932}
\eea
Owing to its nature this operator yields the largest 
enhancement of all the new physics operators. 
Nevertheless it is still safe in respect to experimental bounds. 

The operator $O_1$ with the opposite chiralities, 
$O_1^L\!=\!e m_c \bar c \sigma F (1\!+\!\gamma_5) u$ has a mild suppression by a factor 
$\mhad / m_c$, however it can be enhanced by larger matrix elements
appearing with the right-handed $u$ quark. Therefore we expect to have here
the same numeric estimate as for  $O_1^R$, within a factor of $0.5$ to $2$. 
\vspace*{3pt}

Finally we consider the four-quark operator $O_4$. This is the case of both
the leading-order $1/m_c$ contribution and of the chiral enhancement from the
light valence quarks in the nucleon. Neglecting the strange quarks we have
\beq
O^{(-)}= i\,\Im c_4\,\frac{G_F^2\sin^2{\!\theta_c}\cos^2{\!\theta_c}}{m_c}\,
\left[\bar{d}\gamma^\mu(1\!-\!\gamma_5)d
\:\bar{u}\gamma_\mu\gamma_\nu(1\!-\!\gamma_5)d
\: \bar{d}\gamma^\nu(1\!-\!\gamma_5)u \right] +\mbox{H.c.}
\label{934}
\eeq
The contact photon interaction would come suppressed by two powers of
$1/m_c$. On the other hand, the above leading-$m_c$ interaction enjoys a chiral pion
loop enhancement in the $T$-product with $J_{\rm em}$. Taking the axial 
charge commutator and neglecting the
pole subtraction term in Eq.~(\ref{2030}) the scalar pion vertex 
becomes 
\beq
{\cal G}_s \bar u_p u_n = 
-i\, \Im c_4\,\frac{G_F^2\sin^2{\!\theta_c}\cos^2{\!\theta_c}}{m_c f_\pi}\,
2\matel{p}{\bar{u}\gamma^\mu(1\!-\!\gamma_5)d
\:\bar{d}\gamma_\mu\gamma_\nu(1\!+\!\gamma_5)d
\: \bar{d}\gamma^\nu(1\!-\!\gamma_5)d}{n}_{q =0}.
\label{948}
\eeq
According to our dimensional rules this amounts to 
\beq
|{\cal G}_s|= 
|\Im c_4|\,8 \frac{G_F^2\sin^2{\!\theta_c}\cos^2{\!\theta_c}}{m_c f_\pi}\,
(0.25\GeV)^6 \,\chi_{\rm scal} \, \chi^2_{\rm fl} 
\label{950}
\eeq
and results in
\bea
\nonumber
|d_n|\msp{-7} && \approx 
|\Im c_4|\,\frac{G_F^2\sin^2{\!\theta_c}\cos^2{\!\theta_c}}{\pi^2
  f_\pi^2 m_c}\, g_A \ln{\frac{\mhad^2}{m_\pi^2}}\,
(0.25\GeV)^6 
\,\chi_{\rm scal} \, \chi^2_{\rm fl} \\
&& \msp{25}\approx 6\!\cdot\! 10^{-28}
 e\!\cdot \!\mbox{cm} \cdot \Im c_4 \:\chi_{\rm scal} \,\chi^2_{\rm fl}.
\label{952}
\eea
Using the estimate  Eq.~(\ref{862}) 
we end up with
\beq
|d_n|  \approx 5.7\!\cdot\! 10^{-30\,}
 e\!\cdot \!\mbox{cm} \:\chi_{\rm scal} \,\chi^2_{\rm fl}.
\label{954}
\eeq
Should we apply the simple-minded dimensional estimate to the
$T$-product contribution without considering specifically the pion loop 
or paying attention to the potential chiral enhancement, we would get a
somewhat smaller but a consistent value
\beq
|d_n|  \approx 4\!\cdot\! 10^{-28\,}
 e\!\cdot \!\mbox{cm} \cdot |\Im c_4|\,\chi^2_{\rm fl} \approx 
 4\!\cdot\!10^{-30\,} e\!\cdot \!\mbox{cm} \,\chi^2_{\rm fl}.
\label{956}
\eeq

Therefore, in the case of $O_4$ the induced $d_n$ is about $100$ times the SM. The
origin is evident: $O_4$ has a color structure unfavorable to $D\tto
\pi^+\pi^-$. At the same time, the chirality choice is optimal for both charm
and light quarks, in the nucleon matrix elements. The combination of the
two yields an additional factor of $10$ enhancement in our estimates. 

\begin{table}[h]
\caption{\small The estimated $D^0$ decay amplitudes, the strength of the
CP-odd couplings and expected $d_n$. The two sub-columns for the
chromomagnetic operator $O_2$ correspond to the left-handed (left) 
and right-handed (right) charm fields, respectively.}
  \begin{center}
\begin{tabular}{|c|l|c|c|l|} \hline
 & \hfill $i\matel{\pi^+\pi^-}{O_k}{D^0}$\hfill~~~~~&
 ~$|\sin{\delta_{\scalebox{.5}{FSI}}}\,\Im
 c_k|$~ \rule[3pt]{0pt}{8pt} &  \multicolumn{2}{c|} {$|d_n|,\: e\!\cdot\! \mbox{cm}$} \\ \hline
$O_1$ & ~$8\sqrt{2}\pi \alpha \,q_d \,f_\pi f_+^{D\tto\pi}(0) M_D^2$ & 
~~$5.2 \!\cdot\! 10^{-2}$ \rule[3pt]{0pt}{10pt} & 
 \multicolumn{2}{c|} {$4 \!\cdot\! 10^{-27}$} \\ 
$O_2$ & ~$4\pi g_s\sqrt{3}\,f_\pi f_+^{D\tto\pi}(0) M_D^2$ & 
~~$1.0 \!\cdot\! 10^{-4}$\rule[3pt]{0pt}{10pt} & 
{\small $8 \!\cdot\! 10^{-30}$} & {\small $3 \!\cdot\! 10^{-30}$} \\
$O_3$ & ~$f_\pi f_+^{D\tto \pi}(0) M_D^2$ & 
~~\,$2\;\, \cdot \!10^{-3}$\rule[3pt]{0pt}{10pt} &
\multicolumn{2}{c|} {$\;\;\;    10^{-30}$}  \\
$O_4$ & ~$f_\pi f_+^{D\tto\pi}(0) M_D^2
\;\frac{1}{N_c}\,\frac{\!2m_\pi^2}{\!(m_u\!+\!m_d)m_c\!}$ ~& 
~~$4.6 \!\cdot\! 10^{-3}$ \rule[3pt]{0pt}{10pt} & \multicolumn{2}{c|} {~~$\,10^{-29}$} \\
\hline
\end{tabular}\label{tab}
 \end{center}
\end{table}

For convenience, Table\,\,\ref{tab} summarizes our estimates of $d_n$ in this
section along with the values of $\Im c_k$ from Sect.~\ref{cpdbsm}.

\subsection{A comment on the atomic EDMs}

The atomic size exceeds the nucleon radius by several orders of magnitude, and
as a matter of principle they may have larger EDMs; in
particular, this applies to paramagnetic atoms. The enhanced EDM, however
may originate there mainly through T-violation in the lepton sector, 
with the electron EDM itself or via the induced contact interaction with the 
nucleons. Such manifestations of New Physics are not directly associated with 
the milliweak interaction of quarks and are beyond the subject of the present
study. 

In diamagnetic atoms like mercury the screening mandated by the Schiff theorem
is rather effective and the overall EDM appears to be dominated by the induced
isoscalar CP-odd $\pi^0NN$ coupling affecting non-pointlike electromagnetic
potential of the nucleus  -- yet still at a rather suppressed level,
\beq
d_{\rm Hg} \approx {\cal G}_s \!\cdot\! 3.5 \!\cdot\! 10^{-18\,} e\!\cdot \!\mbox{cm},
\label{1320}
\eeq
see Refs.~\cite{hgtheor,pospritz}. Using, for instance the estimate Eq.~(\ref{950}) we
can expect for the isoscalar coupling $|{\cal G}_s|\!\approx\! 10^{-15}$. 
Therefore, as anticipated  the diamagnetic atom EDMs, while probably not 
yet fully
competitive in sensitivity with the direct $d_n$, may become comparable 
in certain NP scenarios yielding amplitudes with a right-handed light quark, 
owing to  the recent radical improvement \cite{hg199}
in the precision for the $^{199}{\rm Hg}$ EDM.

\section{Conclusions}

The KM mechanism of CP violation in the Standard Model is an instructive
example of a realistic phenomenological theory where the dominant
contribution to the electric dipole moment of neutron comes not from the
effective CP-odd operators of lowest dimension, but via a nontrivial interplay
of different amplitudes at a relatively low energy scale. In the SM this
evidently roots in an intricate nature of the CP violation intimately related
to flavor dynamics requiring existence of a few generations. It may be
interesting to investigate, in general terms, if a similar pattern can naturally
fit theories beyond the SM which would describe
flavor dynamics at a more fundamental level. 

We have argued that in the SM itself with vanishing $\theta$-term 
the neutron electric dipole moment has natural size about 
$10^{-31\,} \textrm{e}\!\cdot\! \textrm{cm}$ and may even exceed this, due to
the interference, at the momentum scale around $1\GeV$ of the two $\Delta
C\!=\!1$, $\Delta S\!=\!0$ weak four-quark amplitudes. This mechanism does not
require short-distance loop effects, is finite in the chiral limit and does
not depend on the strange- vs.\ down-quark mass splitting. 

The CP-odd direct-type $D^0$ decay asymmetry reported recently at the level of
$10^{-2}$ does not naturally conform the expectations in the SM, which are 
typically an order of magnitude smaller. This may be an indication for new CP-odd
forces beyond the SM, although such an interpretation should still be viewed
cautiously. 

If New Physics indeed induce a milliweak CP-odd decay amplitude in charmed
particles, it may also be expected to generate, at the NP scale,
flavor-diagonal CP-odd interactions in the light hadron sector. The EDMs of
nucleons and atoms are extremely sensitive to them, and the existing
experimental bounds place strong constraints on the effective interactions
seen at low energies. Such low-energy effective operators are 
model-dependent and their connection to the charm CP violation is 
indirect, to say the least.

Nevertheless, a certain, possibly subdominant contribution to $d_n$ is
generated at the charm energy scale in a direct analogy with the Standard
Model. It is fully independent of the effects originating from the NP scale
and directly reflects the scale of CP violation in charmed particles. Our
analysis suggests that this would increase $d_n$ compared to the SM prediction
by more than an order of magnitude: the typical enhancement is between $30$ and
$100$, depending on the chiral, color and flavor composition of the charm NP
amplitudes. In an ad hoc case of the CP violation through the 
electromagnetic $c\tto u$ dipole operator alone the neutron EDM can be 
even as larger as 
$5\!\cdot\! 10^{-27\,} \textrm{e}\!\cdot\! \textrm{cm}$. However, the
possibility itself for NP to generate such a CP-odd electromagnetic operator
but not a similar chromomagnetic one of the commensurate strength, does not
look natural. 

We conclude that New Physics CP violation in charm at the reported level
remains safe in respect to existing strong experimental bounds on EDMs, as
long as the direct effects are considered. At the same time it would 
significantly reduce the gap between the bounds and the expected size of the
EDMs, and would make the new generation of the EDM experiments more topical.

In the present analysis we have assumed that a new source of CP violation
appears solely in $|\Delta C|\!=\!1$ interactions. Eventually the known 
flavor dynamics must be embedded in a full picture of flavor together 
with CP violation at some high scale, where new dynamic fields are also 
present. Attempts to investigate the new phenomena along these lines have 
been reported in \cite{gino,gian}, 
considering the observed $\Delta a_{\rm CP}$ e.g.\ in a supersymmetric 
framework. This generically induces additional CP violation compared 
to our scenario, which would modify the impact on the EDMs.

\vspace*{5pt}

\noindent
{\bf Acknowledgments:} 
~We are grateful to I.~Bigi and A.~Khodjamirian for comments. N.U.\ is happy
to thank V.~Petrov for invaluable discussions of the chiral baryon properties,
and acknowledges the hospitality of the HEP theory group of Torino
University where the final part of the paper was written. 
We thank A.~Vainshtein for many discussions and insights.
The work was largely supported by the German research foundation DFG under
contract MA1187/10-1 and by the German Ministry of Research (BMBF), contracts
05H09PSF; it also enjoyed a partial support from the NSF grant PHY-0807959 and
from the grant RSGSS 4801.2012.2.

\section*{Appendix: Generalized Goldberger-Treiman relation and the scalar $\pi NN$ vertex}\label{ggt}
\setcounter{equation}{0}
\renewcommand{\theequation}{A.\arabic{equation}}
\renewcommand{\thetable}{\Alph{table}}
\renewcommand{\thesubsection}{\Alph{subsection}}
\setcounter{section}{0}
\setcounter{table}{0}

CP-odd perturbations in general induce the parity-violating scalar
pion-to-nucleon vertex which may not vanish at small pion momentum. Such
amplitudes often play a special role owing to the small (vanishing in the
chiral limit) pion mass. Here we give a compact current algebra derivation of
the corresponding small-momentum limit for a general CP-odd perturbation
operator $O^-$. We also point out that the induced CP-odd vertex necessarily
vanishes in the chiral limit at zero pion momentum for any operator $O^-$ which does
not involve the right-handed light quarks (their absence may be established 
at arbitrary chosen normalization point).

For simplicity of the notations we consider the charge pion. 
Its amplitude off the nucleon takes the following form in the limit of
vanishing pion momentum:
\beq
A_{\pi^- NN}(0)= \frac{1}{f_\pi} \left[\matel{N}{\mbox{$\frac{1}{i}$}[Q_5^+(0),O^{(-)}]_0}{N} -
\frac{\matel{0}{\mbox{$\frac{1}{i}$}[Q_5^3(0),O^{(-)}]_0}{0}}{-\matel{0}{\bar \psi \psi}{0}}
\matel{N}{\bar{u}d(0)}{N}
\right].
\label{2030}
\eeq
where $Q_5^+\!=\!u^\dagger \gamma_5 d$,  
$Q_5^3\!=\!q^\dagger \frac{\tau^3}{2}\gamma_5 q$ are the axial charge
densities and the equal-time commutators marked with the null subscript 
are calculated according to the standard rules. 

To prove it, we first establish a counterpart of the Goldberger-Treiman
relation for the general case where parity can be violated. To this end we
consider the exact nucleon matrix element of the non-singlet light-flavor axial
current $J_\mu$ (let it be $J_{\mu\,5}^{+}$, for concreteness)
\bea
\nonumber
\matel{N}{J_\mu^5}{N} \msp{-4}& = & \msp{-4} 
g_A(q^2) \bar\Psi_N \gamma_\mu\gamma_5 \Psi_N + b(q^2)
\bar\Psi_N \sigma_{\mu\nu} q^\nu \gamma_5 \Psi_N + C(q^2) q_\mu \bar\Psi_N
\gamma_5\Psi_N  + \\
&  & \msp{-4} 
a(q^2) \bar\Psi_N \gamma_\mu \Psi_N + b(q^2)
\bar\Psi_N \sigma_{\mu\nu} q^\nu \Psi_N +  c(q^2) q_\mu \bar\Psi_N \Psi_N ,
\label{936}
\eea
where the last three terms  violate parity being  induced by
$O^{(-)}$. Consequently, 
\beq
\matel{N}{\partial_\mu J_\mu^5}{N}= (2M_N g_A(q^2)+ q^2C(q^2))\bar\Psi_N
i\gamma_5\Psi_N  + i c(q^2) q^2 \bar\Psi_N \Psi_N.
\label{937}
\eeq
which gives two relations, for the pseudoscalar and for the scalar structures.

Noether theorem relates the divergence of the current obtained from the quark equations
of motion to the variation of the
Lagrangian under the chiral symmetry transformation; in the case of QCD the 
variation
comes from the conventional light quark mass term and from the corresponding
commutator of the axial charge with $O^{(+)}$ which we denote 
by $-i{\cal D}^{+}$:\footnote{For simplicity we assume $m_u\!=\!m_d$.}
\beq
\partial_\mu J_\mu^5 =2 m_q \bar{u}i\gamma_5 d + {\cal D}^{+} , \qquad 
 {\cal D}^{(+)}=  \mbox{$\frac{1}{i}$}[Q_5^+(0),O^{(-)}]_0, 
\label{1937}
\eeq
and therefore 
\beq
\matel{N}{2 m_q \bar{u}i\gamma_5d +{\cal D}^{(+)} }{N}= 
(2M_N g_A(q^2)+ q^2C(q^2))\bar\Psi_N 
i\gamma_5\Psi_N  + i c(q^2) q^2 \bar\Psi_N \Psi_N.
\label{1937a}
\eeq
This equation can be taken in the limit $m_q\tto 0$ and to the first order in
the $O^{(-)}$ perturbation. 
The pseudoscalar structure becomes the Goldberger-Treiman
relation $f_\pi g_{\pi NN}=2M_N g_A$ 
stating the existence of the Goldstine boson through the pole in $C(q^2)$
as long as $M_N g_A(0)\!\ne\!0$. The scalar term dictates that the pion pole
residue likewise carries the scalar nucleon vertex proportional to 
$\matel{N}{{\cal D}^{(+)}}{N}$ at zero momentum transfer:
\beq
A_{\pi NN}=\frac{\matel{N}{{\cal D}^{(+)}(0)}{N}'}{f_\pi}\,.
\label{938}
\eeq
${\cal D}^{(+)}$ in Eq.~(\ref{1937}) is just the conventional 
PCAC commutator. The subtlety is important, however 
that the matrix element above stands for
the {\sl exact} nucleon states rather than for the unperturbed QCD ones as is
usually implied when expanding in perturbation; this fact is indicated by the
prime in Eq.~(\ref{938}). The difference becomes important in the chiral
limit where the pion mass is parametrically small, as illustrated later. 

To bypass this complication we apply a Lagrange multiplier trick, namely consider,
instead, the CP-odd perturbation $O_\lambda^{(-)}\!=\! O^{(-)}\!-\! \lambda
(\bar{u}i\gamma_5 u\!-\!\bar{d}i\gamma_5 d)$ with an arbitrary $\lambda$, and
keep $m_q$ nonzero. 
On one hand, the operator
$\bar{u}i\gamma_5 u\!-\!\bar{d}i\gamma_5 d =\! 1/m_q\,\partial_\mu J_{\mu\,
  5}^{(3)}$ 
is the total derivative in QCD and
does not change any strong amplitude; hence it can be added to the perturbation 
for free. 
On the other hand, $\lambda$ can be
taken such that the perturbation $O_\lambda^{(-)}$ becomes nonsingular in the
chiral limit. (In other words, in this case one can safely perform a double
expansion in $m_q$ and in $O^{(-)}$.) Since the chiral singularity comes from
the pion pole in the correlators, the value of $\lambda$ is determined by
vanishing of the residue $\matel{\pi}{O_\lambda^{(-)}}{0}$, or 
\beq
\lambda = \lim_{m_q\to 0}
\frac{\matel{0}{O^{(-)}}{\pi^0}}
{\matel{0}{\bar{u}i\gamma_5u\!-\!\bar{d}i\gamma_5 d}{\pi^0}} = \lim_{m_q\to 0}
\frac{\matel{0}{\mbox{$\frac{1}{i}$}[Q_5^3(0),O^{(-)}]_0}{0}}{-2\matel{0}{\bar{\psi}\psi}{0}}
\label{2022}
\eeq

With this choice of $O_\lambda^{(-)}$ the exact nucleon states in
Eq.~(\ref{938}) enjoy a regular expansion in both $m_q$ and
in $O_\lambda^{(-)}$ free from a $1/m_q$ enhancement. Therefore, to the first order
in $O_\lambda^{(-)}$ at $m_q\!\ll\!\mhad$ we can ignore the difference
between the exact and the unperturbed nucleon states in Eq.~(\ref{938}). The net
effect of the resummation of the pole terms then amounts to subtracting from 
${\cal D}^{(+)}$ the divergence of the current with the mass term 
$\lambda (\bar{u}i\gamma_5 u\!-\!\bar{d}i\gamma_5 d)$ which is equal to 
$2\lambda\,\bar{u}d$. 
This is the relation Eq.~(\ref{2030}).
\vspace*{3pt}

The presence of a subtraction term beyond the conventional PCAC commutator in
the weak transition amplitudes has been appreciated in the context of
electroweak calculations in the early 1980s \cite{cvvw,pipole}, being evident
when using $\sigma$-models to visualize the chiral symmetry breaking. As noted
in the end of this section, it is likewise intuitive in the ordinary
CP-conserving weak decays, in particular of kaons, without recourse to chiral
Lagrangians. Later it was more systematically incorporated in the chiral
expansion for many EDM calculations beyond the SM \cite{pospelov}.

The above proof, while short and general, may look somewhat 
mysterious since a finite yet calculable 
part of the usual PCAC commutator term appears to be miraculously eaten up 
only as a result of the failure of the conventional chiral expansion. The cover of
mystery is removed once the corresponding diagrams are identified and are accurately
calculated. This is possible using the double expansion, in $O^{(-)}$ and then
in $m_q$. We illustrate this in what follows.

\begin{figure}[hhh]
\vspace*{2.5pt}
 \begin{center}
\includegraphics[width=16.cm]{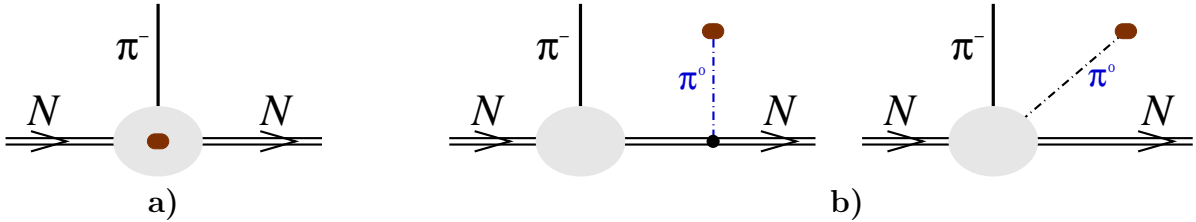}\begin{picture}
(0,0)(0,0)\put(-140,-4){{\bf a)}} \put(-49.3,-4){\bf {b)}}\end{picture}
\vspace*{-15pt}
 \end{center}
\caption{ \small
Examples of contact {\bf a)} and pole {\bf b)} diagrams for the induced $\pi^- NN$
vertex. The pole diagrams may contain non-singular terms as well and these are
included in the contact part of the amplitude. The light shaded blob
represents the pion amplitudes off the nucleon, the solid block shows 
the insertion of the CP-odd operator $O^{(-)}$. The dashed-dotted line is the $\pi^0$
propagator with an infinitesimal momentum.
 }
\label{tomwein}
\end{figure}

The pion-nucleon amplitude to the first order in perturbation $O^{(-)}$ has
two pieces given by the irreducible and the pole diagrams, respectively, see
Fig.~\ref{tomwein}. The latter are those which become singular in the limit of
$m_q\tto 0$ or at vanishing pion momenta. We need to consider them in the
kinematics where $\pi^-$ has a finite momentum yet small compared to the
hadronic scale $\mhad$, while $\pi^0$ has nearly vanishing momentum driven down
by small $m_q$. The pole diagrams in Fig.~\ref{tomwein}b have an enhancement
$1/m_q$ from the pion propagator at zero momentum. The Adler consistency
condition guarantees that the $1/m_q$ enhancement in the pole diagrams is
canceled, but it does not protect against the finite piece we are interested
in.

\thispagestyle{plain}

The conventional PCAC vertex derived from the axial charge commutator, the first
term in Eq.~(\ref{2030}), is just the above contact vertex. 
To determine the extra finite part we take $O^{(-)}\!=\!\bar{u}i\gamma_5
u\!-\!\bar{d}i\gamma_5 d$. Using the operator identity 
$$
\bar{u}i\gamma_5u\!-\!\bar{d}i\gamma_5 d = \frac{1}{m_q} \partial_\mu J_{\mu\,
  5}^{3}
$$
we have 
\beq
\matel{N\pi}{(\bar{u}i\gamma_5 u\!-\!\bar{d}i\gamma_5 d)(0)}{N} = 
i\frac{q_\mu}{m_q} \matel{N\pi}{J^{3}_{\mu\,5}(0)}{N}=0 \mbox{~at~} q_\mu\tto 0.
\label{2020}
\eeq
This equation is valid for {\sl arbitrary} (even large!) nonzero $m_q$ and 
arbitrary $\pi^-$ momentum.

We can examine it to the first two orders in $m_q$. Since there is a pion
propagator pole, the leading constraint is the vanishing of the $1/m_q$ piece.
This is the Adler consistency condition: the pole residue proportional to the 
$\pi^0$ emission amplitude at zero momentum vanishes. 
Vanishing of the ${\cal O}(m_q^0)$
term means the cancellation of the two contributions, the pole and the contact
amplitudes; no further terms appear due to the Adler condition established at
the previous step. 
The latter comes from various regular terms not containing
poles or kinematic singularities and therefore can be calculated simply 
at $m_q\!=\!0$ and $p_{\pi^-}\tto 0$. The standard PCAC commutator applicable 
for soft $\pi^-$ is just this piece.  The former contribution is the new 
chirality-violating terms
$\propto \!m_q$ which spoil the exact Adler cancellation of the amplitude in
the chiral limit. Eq.~(\ref{2020}) fixes it to be exactly minus the contact
amplitude:
\beq
\matel{0}{(\bar{u}i\gamma_5 u\!-\!\bar{d}i\gamma_5 d)(0)}{\pi^0} \cdot 
\frac{A(N\pi^-\to N\pi^0)}{m_\pi^2} = 
-\matel{N\pi^-}{(\bar{u}i\gamma_5 u\!-\!\bar{d}i\gamma_5 d)(0)}{N}.
\label{2024}
\eeq
This relation is exact in the limit of $p_{\pi^{0}}\tto 0$ and is similar 
in spirit to the Tomozawa-Weinberg formula \cite{tomozawa}, 
yet is simpler and differs since
only one pion is soft. It actually applies to any hadron state, not only
$N\pi^-$.

Now we can go back to the case of a general $O^{(-)}$. 
It cannot anymore be represented as a total derivative, and the matrix element 
$\matel{N\pi^-}{O^{(-)}}{N}$ does not vanish. It is still given by the sum of
the contact and the pole diagrams. The former would again be given, for soft 
$\pi^-$, by the Goldberger-Treiman
commutator; it should be taken over the unperturbed nucleons, since
the quark masses are kept nonzero. The latter, the chirally enhanced pole 
diagrams with the strong vertices corrected at order $m_q$ depend only on
$m_q$ but not on $O^{(-)}$, i.e.\ they are given by QCD proper. 
The CP-odd operator $O^{(-)}$ enters them
only at the tadpole $\matel{0}{O^{(-)}(0)}{\pi^0}$, Figs.~\ref{tomwein}b. 
Multiplying the tadpole by the strong
amplitude $A(N\pi^-\to N\pi^0)$ over $m_\pi^2$ from Eq.~(\ref{2024}) we get,
for the CP-odd part,  
\beq
\matel{N\pi^-}{O^{(-)}(0))}{N} - 
\frac{\matel{0}{O^{(-)}(0)}{\pi^0}}
{\matel{0}{(\bar{u}i\gamma_5 u\!-\!\bar{d}i\gamma_5 d)(0)}{\pi^0}} \cdot 
\matel{N\pi^-}{(\bar{u}i\gamma_5 u\!-\!\bar{d}i\gamma_5 d)(0)}{N}.
\label{2028}
\eeq
This representation has an advantage of still being valid at arbitrary $\pi^-$
momentum, yet only to the leading order in $m_q$ (which here means discarding
$m_q/\mhad$). It clearly conforms to Eq.~(\ref{2022}).

The explicit form of the $N\tto N\pi$ amplitudes at arbitrary pion momentum is
not known, therefore to have a concrete expression we finally should assume
$p_{\pi^-} \!\ll \! \mhad$. Using the generalized Goldberger-Treiman  
relation for the conventional
unperturbed nucleon states ($m_q$ is finite now) we expectedly arrive at
Eq.~(\ref{2030}). Thus, we have traced how the chiral pole resummation 
generates the subtraction term exactly in the way anticipated in
our original simple derivation.  \vspace*{3pt}

The additional general observation is useful in view of the left-handed
structure of the weak currents in the SM. Namely, for any operator not containing
right-handed  $u$- or $d$-quark fields the induced $\pi NN$ coupling at zero
pion momentum must vanish in the chiral limit $m_{u,d}\tto 0$. In the cases
where the commutator with the axial charge does not vanish explicitly this implies the
vanishing of the corresponding zero-momentum matrix element. This applies to 
any on-shell amplitude off the hadrons, not only to the nucleon vertex, and is
a counterpart of the Adler consistency condition. 

The reason is that at  $m_{u,d}\!=\! 0$ the theory is invariant under the
isotriplet right-handed chiral transformation 
$$
q(x) \tto e^{i\frac{\alpha}{2}(1\!+\!\gamma_5)\tau^3} q(x);
$$
as long as ${\cal L}_w$ is free from $u_R$ and $d_R$ this symmetry persists in
the full theory including the weak interaction. Since it is spontaneously broken
by the conventional strong dynamics, there is an exactly massless (at
$m_q\!=\! 0$) Goldstone
boson, $\pi^0$, associated with the corresponding exactly conserved Noether
current. This current is evidently a sum of the usually considered axial
current and of a flavor-diagonal vector current. Likewise, as in the conventional
axial-current case, the corresponding Goldstone vertex off the exact
eigenstates vanishes at zero momentum. 

The formal derivation is straightforward if one considers, instead of the conventional
axial current, the corresponding left-handed current. Its divergence, the
analogue of Eq.~(\ref{1937}) vanishes at $m_q\!=\! 0$ by virtue of the exact 
quark field equations of motion, and the generalized Goldberger-Treiman 
relation Eq.~(\ref{1937a}) 
says that $c(0)\!=\!0$, cf.\ Eq.~(\ref{938}). The existence itself of the 
exact Goldstone boson follows from Eq.~(\ref{1937a}) considered in the limit
of small nonvanishing $m_q$ with ${\cal D}\!=\!0$.

Therefore, any weak pion amplitude vanishes for small pion momentum in the
chiral limit unless weak Lagrangian contains right-handed $u$ or $d$ fields.
More generally, it may only remain finite if there is no combination of vector and
non-anomalous axial transformation that leaves ${\cal L}_w$
invariant. Furthermore, the invariance can be checked at arbitrary normalization
scale, and usually it is most evident for the bare operators. As an example,
the bare ${\cal L}_w$ in the Standard Model contains only left-handed fields, 
but Penguins \cite{penguins}
induce the operators with the right-handed light quarks in the conventionally 
considered effective renormalized Lagrangian. Nevertheless the zero-momentum
pion amplitude vanishes in the chiral limit in the Standard Model.

Unlike the pseudoscalar vertex, the induced CP-violating scalar pion-nucleon vertex
describes the Lorentz structure in the amplitude that does not vanish at zero
momentum transfer. Therefore, it must vanish in the chiral limit unless the 
weak interactions include right-handed $u$ or $d$ fields.
\vspace*{3pt}

Concluding the brief discussion of the application of the current algebra
technique we note that the similar methods can be applied, for instance to the
usual weak decays, e.g.\ of kaons or hyperons, both parity-conserving and
parity-violating. In the parity-conserving $\Delta S\!=\!1$ decays we may
subtract from the weak Lagrangian the scalar operator $\bar{s} d$ with an
arbitrary coefficient $\lambda$, rewriting it as 
$\partial_\mu (\bar{s}\gamma_\mu d)/(m_s\!-\!m_d)$. This demonstrates that the
decay amplitude is not changed regardless of $\lambda$.  For parity-violating
transitions we can subtract $\bar{s}i\gamma_5 d\!=\!\partial_\mu 
(\bar{s}\gamma_\mu\gamma_5 d)/(m_s\!+\!m_d)$. This is also useful in establishing
the absence of the chiral enhancement in the $K$-decay amplitudes mediated by composite
quark bilinears like $\bar{s} i\sigma_{\mu\nu} G^{\mu\nu}d$, and to elucidate
other similar cancellations. We do not expand on the related applications 
here.

\end{document}